\documentclass[letterpaper,twocolumn,10pt]{article}
\usepackage{usenix2019_v3}

\usepackage{tikz}
\usepackage{amsmath}

\usepackage{cite}
\usepackage{amsmath,amssymb,amsfonts}
\usepackage{algorithmic}
\usepackage{graphicx}
\usepackage{textcomp}
\usepackage{xcolor}
\usepackage{comment}
\usepackage{cuted}

\definecolor{mypink}{rgb}{0.858, 0.188, 0.478}
\usepackage{xparse}
\usepackage{graphicx}
\usepackage{subfig}
\usepackage[most]{tcolorbox}
\usepackage{amsmath, amssymb, amsthm}
\usepackage{booktabs}
\usepackage{enumitem}
\usepackage{hyperref}
\usepackage{xcolor}

\usepackage{algorithm}
\newcounter{packednmbr}

\newcommand{\oops}[1]{*} 

\newcommand{\todo}[1]{\textcolor{red}{\emph{#1}}}

\newcommand{\pw}{\mbox{Perf/Watt}}

\newcommand{\lpiga}[1]{{\color{red}[Leo: #1]}}

\newcommand{\pmin}{p_{\min}}
\newcommand{\pmax}{p_{\max}}
\newcommand{\Ptot}{P_{\mathrm{total}}}

\newcommand{\Pnet}{P_{\mathrm{net}}}

\begin{document}
%-------------------------------------------------------------------------------

%don't want date printed
\date{}

\newcommand{\iscasubmissionnumber}{50}
%%%%%%%%%%%%%%%%%%%%%%%%%%%%%%%%%%%%

\pagenumbering{arabic}
%%%%%%%%%%%---SETME-----%%%%%%%%%%%%%
%\title{
%End-to-End Power Management for a 150 MW AI Datacenter: \\ From Planning to Runtime Optimization\vspace{-5mm}}
\title{
Provisioning to Runtime Optimization of a 100 MW-Scale AI Cluster}

\author{Ehsan K. Ardestani, Leonardo Piga, Jovan Stojkovic, Pavan Balaji, Mustafa Ozdal, \\
Mikel Jimenez Fernandez, 
Mihaela Dimovska,
Luka Tadic,
Hao Shen,
Devika Vishwanath,\\
Richa Mishra,
Melaku Mihret,
Valentin Andrei,
Mauricio Cespedes,
Julien Prigent,
James Monahan,\\
Tyler Graf,
Bin Li,
Shobhit Kanaujia,
Kaushik Veeraraghavan,
Chunqiang Tang\\
\\
Meta Platforms}

\maketitle

\begin{abstract}
The electric power supply for AI datacenters is now the most significant bottleneck in the race toward Artificial General Intelligence (AGI), surpassing even the constraint of AI accelerator availability. To our knowledge, this paper is the first to describe the end-to-end power management process for a hyperscale AI datacenter—from early power planning to accommodate next-generation accelerators 6–12 months before their general availability, to tuning power settings after large-scale deployment, and finally to dynamic, runtime power management for evolving workloads. We present detailed power measurements for a 150 MW datacenter hosting a cluster of 83K GB200 GPUs connecting through RDMA back end network. We share insights from building this state-of-the-art AI cluster. We hope this work encourages practitioners across the industry to share their own experiences as well.
\end{abstract}

%Contributions
%	- perf/watt (performance/power) - projection provision
%	- details of power delivery
%		- special and temporal locality
%	- Dynamic TDP @ DC scale
%		- Data for Dimmer
%			- From an MSB that is high loaded
%	- Smoother {SW, HW}

\section{Introduction}

The electric power supply for AI datacenters has overtaken the supply of AI accelerators as the primary limiting factor in the development of Artificial General Intelligence (AGI). 
%For example, Meta has announced plans to build multiple datacenters with gigawatts (GW) of compute each~\cite{hyperion}.
For example, OpenAI has announced plans to build tens of gigawatts (GW) of compute with various partners~\cite{openai-stargate,openai-nvidia-10gw,openai-amd-6gw,openai-broadcom-10gw}, 
%and it has even been reported to be targeting 250 GW of compute by 2033~\cite{openai-250gw}. 
For perspective, total U.S. power-generation capacity increased from 1,189 GW in 2023 to 1,230 GW in 2024—an addition of only 41 GW~\cite{power-2023,power-2024}. This indicates that AI datacenters would consume a significant portion of the new power capacity being added in the United States.

Given the importance of power, numerous studies have explored device and datacenter power management~\cite{power-survey-1,power-survey-2,power-survey-3}. Most focus on lowering utility costs, optimizing the use of green energy, or exploiting power oversubscription and capping to maximize power utilization. 

However, these studies overlook two real-world issues. First, because power is a scarce resource and datacenters operate under a fixed maximum utility power supply, the primary goal of datacenter power planning, particularly in the AGI race, is to maximize the total compute capacity of AI accelerators that can fit within the power budget. Although many studies focus on reducing electricity costs, this is secondary: datacenter and accelerator expenses far exceed the utility bill. Moreover, failing to fully utilize available power leads to a shortage of compute capacity, incurring an even higher opportunity cost.

Another limitation of prior studies is their narrow focus on only the final phase of the three-phase power management process: (1) \textbf{datacenter-planning phase}: operators often lack direct access to the next generation of accelerators (e.g., NVIDIA VR200), and must make decisions 6--12 months in advance using only limited performance and power data from vendors. As a result, they need to estimate the ideal power settings for accelerators, as well as power allocation for networking gears, and decide how many to purchase, and start deploying the necessary network and cooling infrastructure; (2) \textbf{deployment-validation phase}: after deployment, refining accelerator power settings based on measured consumption under representative workloads; and (3)~\textbf{operational phase}: dynamically tuning power settings during the execution of uncontrolled, live workloads to maximize performance within the power budget. Unlike prior work, which primarily addresses phase 3, we believe this is the first to present end-to-end production experience across all three phases of power management for a hyperscale AI cluster; specifically, a 150-megawatt (MW) cluster consisting of 83K GB200 GPUs.

Our contributions in this paper are not only to share real-world experience in end-to-end power planning and management, but also to report a set of interesting or surprising findings that can help guide future practitioners. 

First, in the datacenter-planning phase, we do not use the accelerator’s Thermal Design Power (TDP) to determine the maximum number of accelerators that can fit within the datacenter's power budget. Instead, we assume a lower power limit based on the accelerator's projected power-performance curve, minimizing performance loss while maximizing the number of accelerators and overall compute capacity. This has led the industry to adopt offerings with different power limits products or operation points to signify cluster throughput vs max GPU performance. %For GB200, this power limit is set to $\approx 80\%$ of its TDP. At runtime, when rack components consume less power than assumed in planning, we use the residual power to boost accelerator performance by increasing operating power beyond this 80\% limit.

Second, different sub-areas of a datacenter contain a complex, uneven mix of hardware, such as GPU-server racks, Network racks, CPU-server and storage racks, and cooling devices. Even with identical, static workloads on similar racks, hardware heterogeneity causes considerable power imbalances across different power delivery paths, limiting dynamic power management. For example, increasing the power limit for all accelerators in a large training job is constrained by the power delivery path with the least headroom.

Third, accelerator performance does not scale uniformly with the power limit; the relationship is complex. For example, GB200's HBM bandwidth stays constant as the power limit decreases from 1200 Watt~(W) to 1000 W, but drops sharply by 15\% when further reduced to 800 W. Additionally, matrix multiplication performance is affected not only by the power limit but also by the operation's arithmetic intensity, which is workload dependent.

Fourth, in large-scale synchronous AI training, allowing individual accelerators and racks to manage power settings locally is ineffective, as stragglers slow down the entire job. Instead, when a power-delivery device nears its limit, we lower the power limits for all accelerators it governs, ensuring consistent and collaborative performance reduction.

Fifth, we do not directly use the Power Supply Unit (PSU)’s rack power measurement, as it consistently overestimates power usage to ensure safe operation within the power budget. Instead, we rely on sensors at the reactor power panel (RPP) level, which more accurately report aggregate power consumption across multiple racks. We use this data to calibrate and adjust the PSU readings.

Finally, to maximize compute capacity within a datacenter's power budget, it is often best to choose the latest accelerators. For example, compared with  H100, GB200 delivers twice the performance per watt, thanks to its higher FLOPS, greater network bandwidth, and larger scale-up domains.

In summary, by sharing these insights and our production experience with a 150~MW AI cluster, we hope to inspire practitioners across the industry to share their own experiences as well.

\section{Hosting AI Workloads}

\subsection{AI Workloads and Power-Performance Trade-offs}

% \vspace{2pt}

\noindent \textbf{Workloads.} Given the cluster can be used for either training or inference jobs, of the three power management phases, the first two (planning and deployment validation) must assume the most power-hungry workload, which is pre-training, as it requires the largest scale of accelerators per job and presenting the most stringent power demands at the datacenter level. The operational phase will behave differently for inference vs pre-training. Hence in this work, we focus on pre-training and consider a range of Mixture of Expert Models.

\begin{comment}
    
AI workloads continue to evolve rapidly with advances in algorithms, data, and compute capabilities. At a high level, current workloads can be categorized into \emph{pre-training}, \emph{post-training}, and \emph{inference}. 

Pre-training develops foundational models by processing trillions of tokens, typically requiring the largest scale of accelerators and presenting the most stringent power demands at the datacenter level. Post-training adapts base models through supervised fine-tuning or reinforcement learning~\cite{ouyang2022training,du2025post}, while inference encompasses both online serving and offline synthetic data generation~\cite{chen2024large}. 

For this paper, 
% among all the use cases, 
we focus on pre-training workloads, although the cluster use is not limited to pre-training. This is because pre-training represents the most challenging scenario for power and performance provisioning and optimization at scale.

\end{comment}

\noindent \textbf{Bottleneck resources.} 
The execution of large-scale AI workloads can be decomposed into distinct operations: \emph{compute-bound}, \emph{memory-bound}, and \emph{communication-bound}. Compute-bound operations are dominated by intensive arithmetic tasks, typically leveraging specialized hardware (e.g. Tensor Cores), and are power bound. 
Memory-bound operations are limited by the rate at which data can be moved between high-bandwidth memory (HBM) and the compute units. HBM bandwidth in GB200 genration is not power bound in the power limit range if interest. 
Communication-bound operations involve data exchange across nodes, often over high-speed interconnects, and are integral part of distributed training. Typically these phase of operation is not power bound.

\subsection{AI Cluster Architecture and Performance Metric}
\label{sec:cluster_arch}

\noindent \textbf{Datacenter.} The AI datadenter is one of our company's traditional data centers, featuring air-cooled hardware and no facility based liquid cooling. The region consists of 5 buildings, each with around 30~MW. Each building contains 4 data halls, which are failure domains with 3 Main Switch Board (MSB) each. This 150 MW AI datacenter is part of the larger 1~GW AI cluster buildout. 
%NAO is part of the larger 1~GW \textit{Prometheus} cluster~\cite{metagw}.
% , enabling 1~GW of IT power.

% \vspace{2pt}
\noindent \textbf{Cluster Composition.} An AI cluster is an integrated system comprising several interdependent hardware subsystems connected via a dual-network architecture. 
First, 
a high-throughput, low-latency back-end network based on RDMA provides 800~Gbps 
per GB200 GPU for internode communication among GPUs. In addition, a front-end TCP/IP Ethernet offers 200~Gbps per GPU for general connectivity. 

Besides GPU servers, the datacenter includes various supporting services, such as high-performance storage for massive datasets, CPU servers for data processing, and the network fabric required to deliver data to accelerators at line rate.

\noindent \textbf{Channel Island Catalina Rack (GB200).} The core of an AI cluster are accelerator-enabled compute racks, which for this study is powered by the Catalina pod~\cite{catalinaocp}. 

\begin{figure}[h]
    \vspace{-1mm}
    \centering
    \includegraphics[width=1\linewidth]{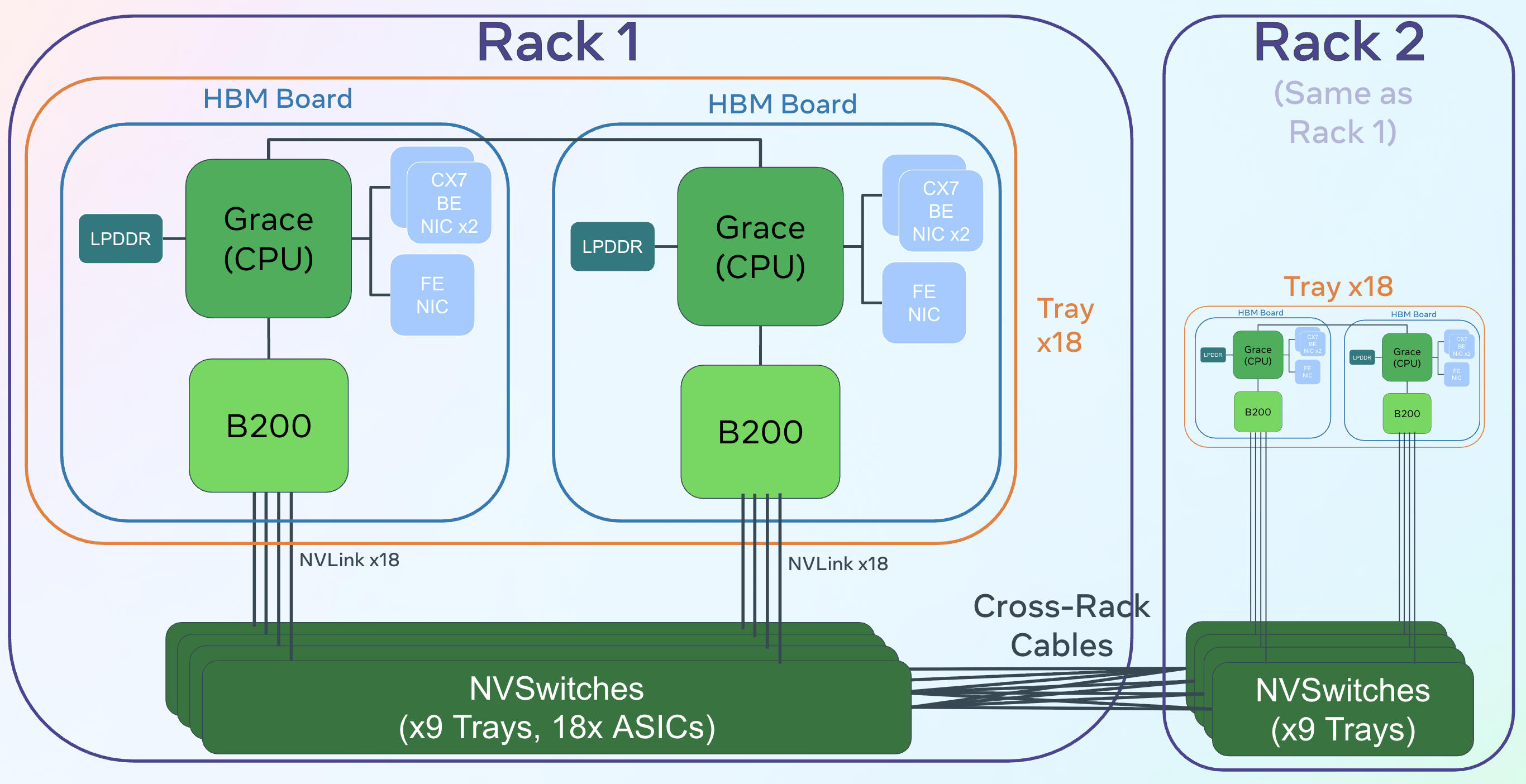}
    \caption{The Catalina pod GB200 configuration.}
    % 2 Back End CX7 attached to the CPU provide 100 GB/s of RDMA Network bandwidth per GPU}
    \label{fig:catalina}
    \vspace{-2mm}
\end{figure}

The Catalina pod architecture is different from NVIDIA's GB200 reference design, as it contains 2 IT racks each hosting 36 GB200 GPUs. The racks are connected through trunk cables connecting NVSwitches in each rack to form a 72-GPU NVLink scale up domain (2x36). The primary reason for this specific configuration is to enable 100~GB/s of scale out RDMA bandwidth per GPU, instead of 50~GB/s in the reference design. As shown in Figure~\ref{fig:catalina}, 
this is made possible by connecting two 400~Gbps CX7 NIC per Grace CPU for the backend network. 

\begin{figure}[h]
    \vspace{-1mm}
    \centering
    \includegraphics[width=1\linewidth]{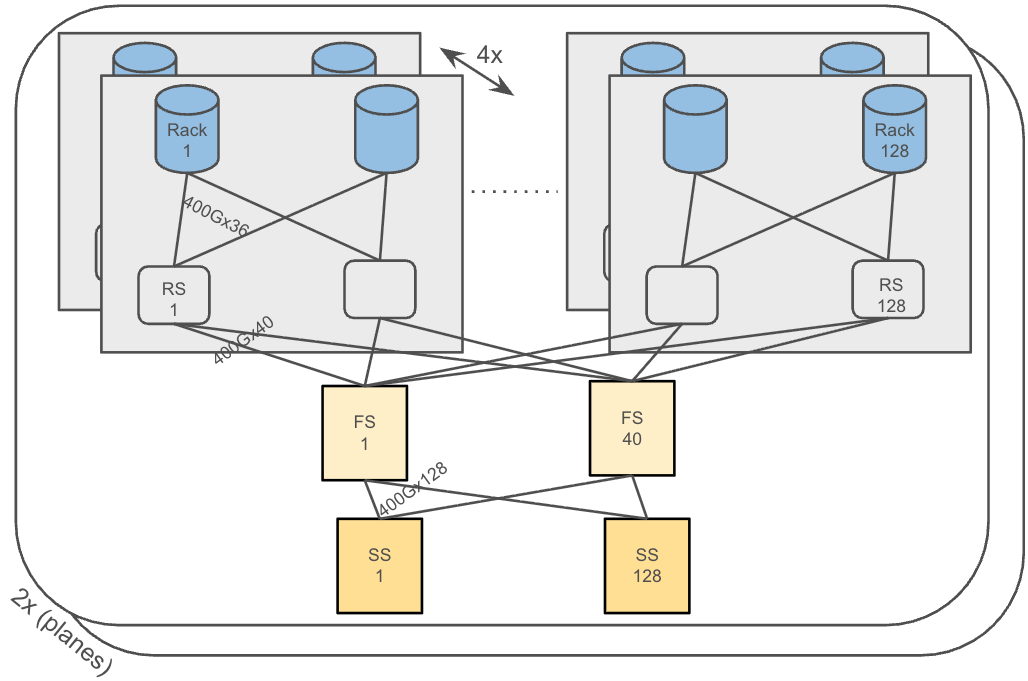}
    \caption{Back end network schematics. There are 3 levels of switches, Rack Swithc (RS), Fabric Switch (FS) and Spine Switch (SS). Each CX7 NIC is forming a plane Packets are sprayed after RS to improve entropy in the network.}
    \label{fig:be_topology}
    \vspace{-2mm}
\end{figure}

\begin{figure}[!htb]
\vspace{0mm}
    \centering
    \includegraphics[width=\linewidth]{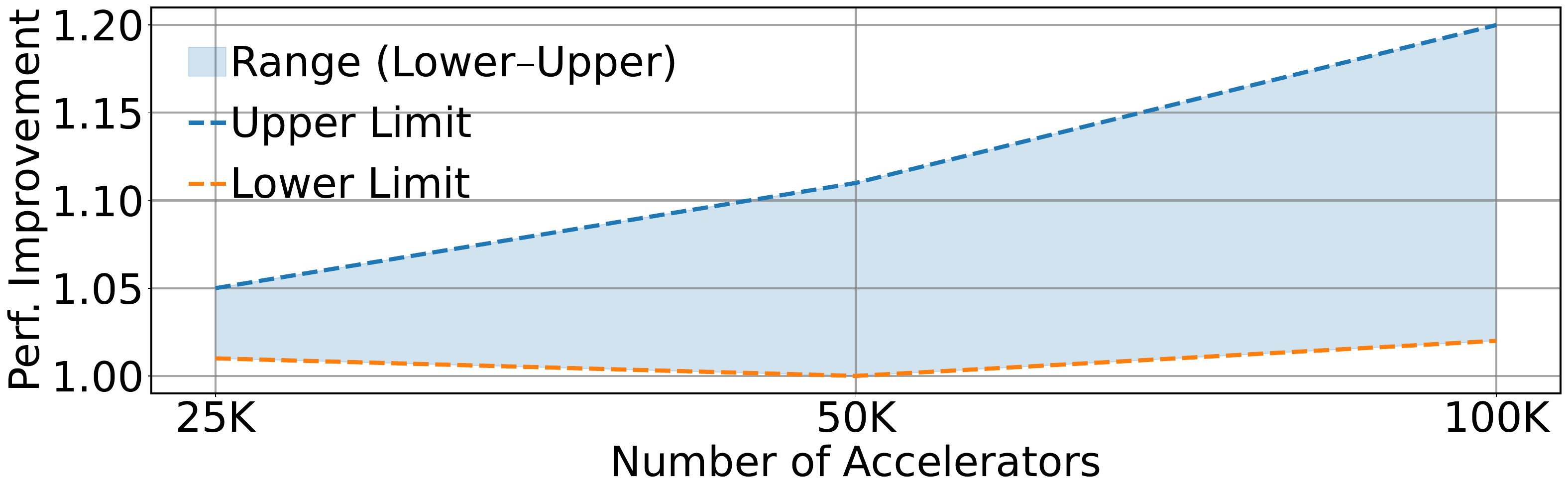}
    \caption{Performance improvement of a system with 100GB/s per-GPU bandwidth over a system with 50GB/s per-GPU bandwidth while varying the cluster size.} 
    \label{fig:gb200_scaleout_bw}
    \vspace{-2mm}
\end{figure}

Due to limited IO per Grace CPU, this is only possible by using one CPU per GPU, resulting in a host with 2+2 GPU+CPU instead of 4+2 in the reference design.  Figure~\ref{fig:gb200_scaleout_bw} shows the projected impact of 100~GB/s vs 50~GB/s scale out bandwidth per GPU. Given we expect this cluster to be part of the larger 1~GW buildout, %\textit{Prometheus} cluster~\cite{metagw}, 
we optimized this platform for 100~GB/s capability per GPU. The back end network is comprised of 3 levels of switches as shown in Figure~\ref{fig:be_topology}.

\begin{figure}[h]
    \vspace{-3mm}
    \centering
    \includegraphics[width=1\linewidth]{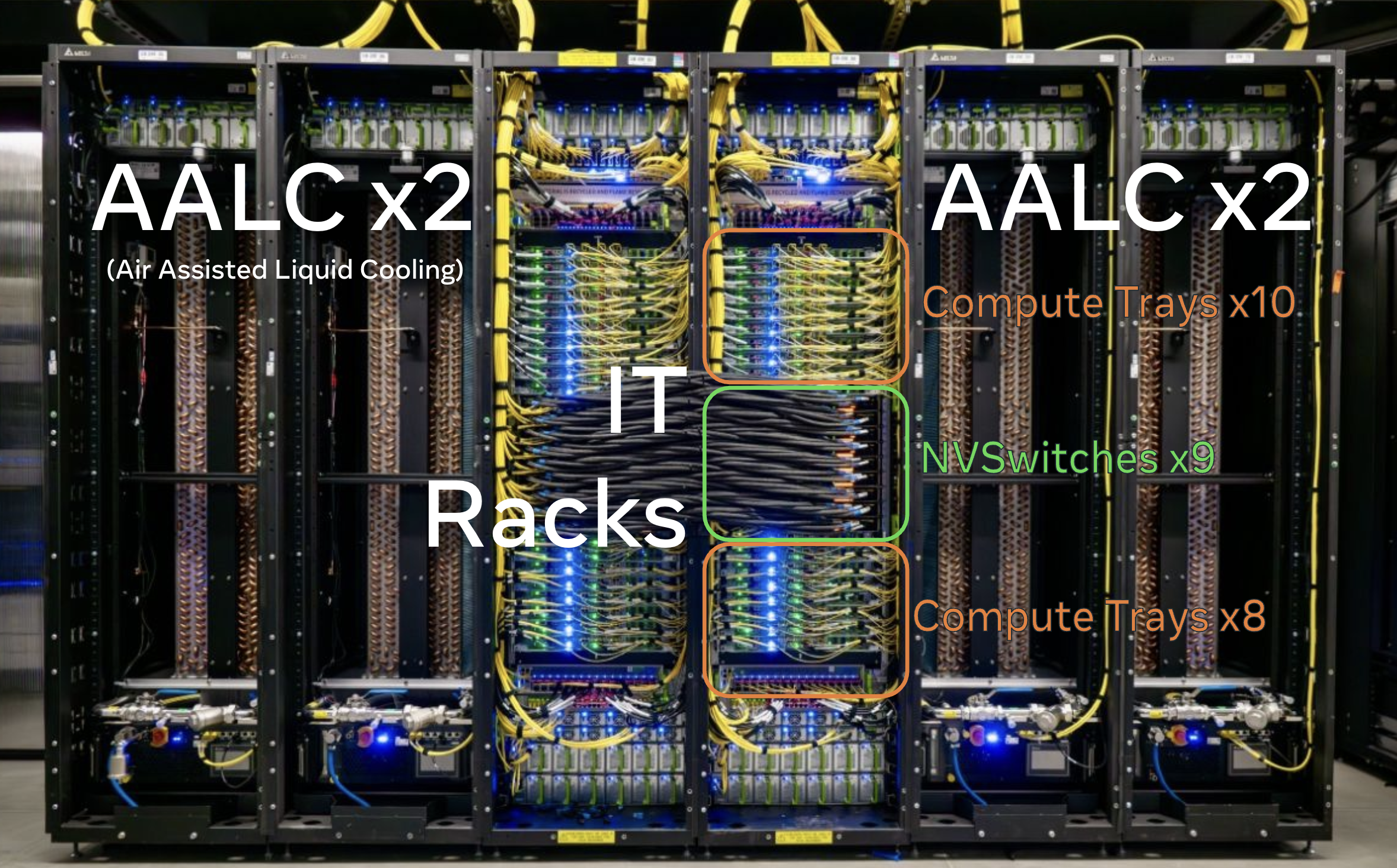}
    \caption{A Catalina-based GB200 pod with  two interconnected IT racks and two AALC per rack, hosting 72 GB200 GPUs.}
    \label{fig:cic-gb200}
    \vspace{0mm}
\end{figure}

Figure~\ref{fig:cic-gb200} shows the composition of a Catalina pod.
GB200's power limit of up to 1.2~kW per accelerator necessitates liquid cooling to maintain safe and efficient operation. In a Catalina pod, each compute rack is paired with two Air Assisted Liquid Cooling (AALC) racks to manage the thermal load generated by high-density compute in the absence of facility-provided chilled water.  
 
%These AALC systems circulate coolant through cold plates attached to the accelerators, transferring heat to a secondary air loop that is then expelled from the data hall.  This approach adds additional power overhead for pumps, fans, and heat exchangers, which must be factored into the overall power provisioning, particularly during periods of elevated ambient temperature, when the efficiency of heat rejection systems may be reduced and the cooling load increases.

\begin{comment}
    
\vspace{2pt}
\noindent \textbf{Holistic Provisioning.} The process of data center provisioning and any power optimization framework must holistically address all cluster subsystems: 
compute, networking, storage, and cooling. 
Insufficient provisioning of any single component can introduce critical perform/ance bottlenecks, leading to under-utilization of accelerator hardware. 
Such inefficiencies directly reduce computational throughput and degrade the overall efficiency of the deployment. 
Therefore, a holistic design and provisioning approach is imperative to achieve both peak performance and optimal power utilization at scale.

\end{comment}

\vspace{2pt}
\noindent \textbf{Holistic Provisioning.}
Datacenter provisioning for large-scale AI clusters requires a holistic, balanced approach across compute, networking, storage, and cooling, ensuring no single subsystem becomes a bottleneck. 
For instance, insufficient cooling can cause thermal throttling that limits computational throughput, while inadequate network bandwidth or a suboptimal topology can create communication bottlenecks that stall distributed training and reduce hardware utilization. 
We therefore explicitly model and coordinate the provisioning of all critical subsystems to maximize accelerator utilization, maintain predictable performance, and achieve high efficiency at scale. 
To make this concrete, we define cluster performance as the aggregate throughput of large-scale pre-training that uses all available GPUs, assuming a well-engineered training software stack that effectively overlaps computation and communication~\cite{llama3-parallelism} and robustly handles GPU failures, allowing performance to scale with the number of GPUs.

\section{Power for AI Infrastructures}

\subsection{Power Delivery Hierarchy and Constraints}
\label{sec:powerhierarchy}

\begin{figure}[h]
\vspace{-1mm}
    \centering
    \includegraphics[width=\linewidth]{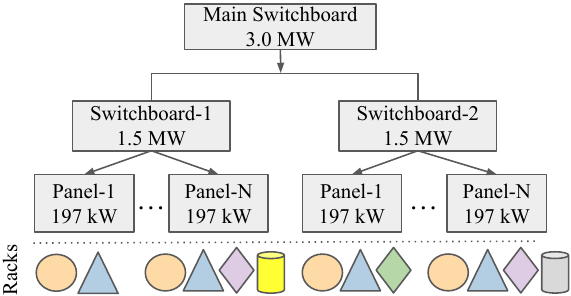}
    \caption{Power distribution hierarchy. Each shape is a different rack type, highlighting the heterogeneous hardware mix and, thus, the varying power headroom across power domains.}
    \label{fig:power_delivery}
    \vspace{-3mm}
\end{figure}

\vspace{2pt}
\noindent \textbf{Power Distribution Architecture.}
In the large-scale AI datacenter, power is delivered through a hierarchical network of electrical devices, each with its own capacity limits. 
At the top is the main switchboard (MSB), which receives power from the utility grid after voltage conversion. 
Below the MSB are intermediate switchboards (SBs), and at the lowest level are reactor power panels (RPPs), which directly supply power to the racks. 
Figure~\ref{fig:power_delivery} shows an example of such a hierarchy.

\vspace{2pt}
\noindent \textbf{Capacity Planning and Over-Subscription.}
Each device in the hierarchy sets an upper bound on the power that can be drawn by all downstream equipment.
For example, each RPP can be rated for up to 197.5~kW. In contrast, a typical MSB may be rated for up to around 3~MW, which must be allocated between IT equipment and mechanical systems like cooling. For planning purposes, we assumed 2.7~MW as for planned power budget to IT equipment and 300~kW for mechanical systems. In practice, these limits are 
% not strictly enforced and do
not 
% represent 
hard constraints and are only useful for power planning. Historically, our DC deployments had assumed 2.5~MW of IT power per MSB.

In our infrastructure, datacenters often \emph{over-subscribe} power at lower levels of the hierarchy, meaning the sum of the rated capacities of all panels or racks beneath a switchboard can exceed the switchboard’s own rating. 
This approach provides flexibility in rack placement and workload scheduling, but it requires careful management: the total power allocated to all racks under any given device must never exceed that device’s rated limit. 
Rack placement algorithms and power management systems must enforce these constraints to avoid overloading any part of the power delivery infrastructure.

\vspace{-3mm}
\subsection{Mechanical Power and Cooling Demand}

\noindent \textbf{Cooling Power Dynamics.}
A portion of the datacenter's power budget is consumed by mechanical systems, primarily for cooling. The power required for cooling varies with environmental conditions, especially ambient temperature. In hotter periods, more power is needed to operate chillers, pumps, fans, and other  equipment to maintain safe temperatures. % for the IT hardware.

\vspace{2pt}

\begin{figure}[h]
    \centering
    \includegraphics[width=0.9\linewidth]{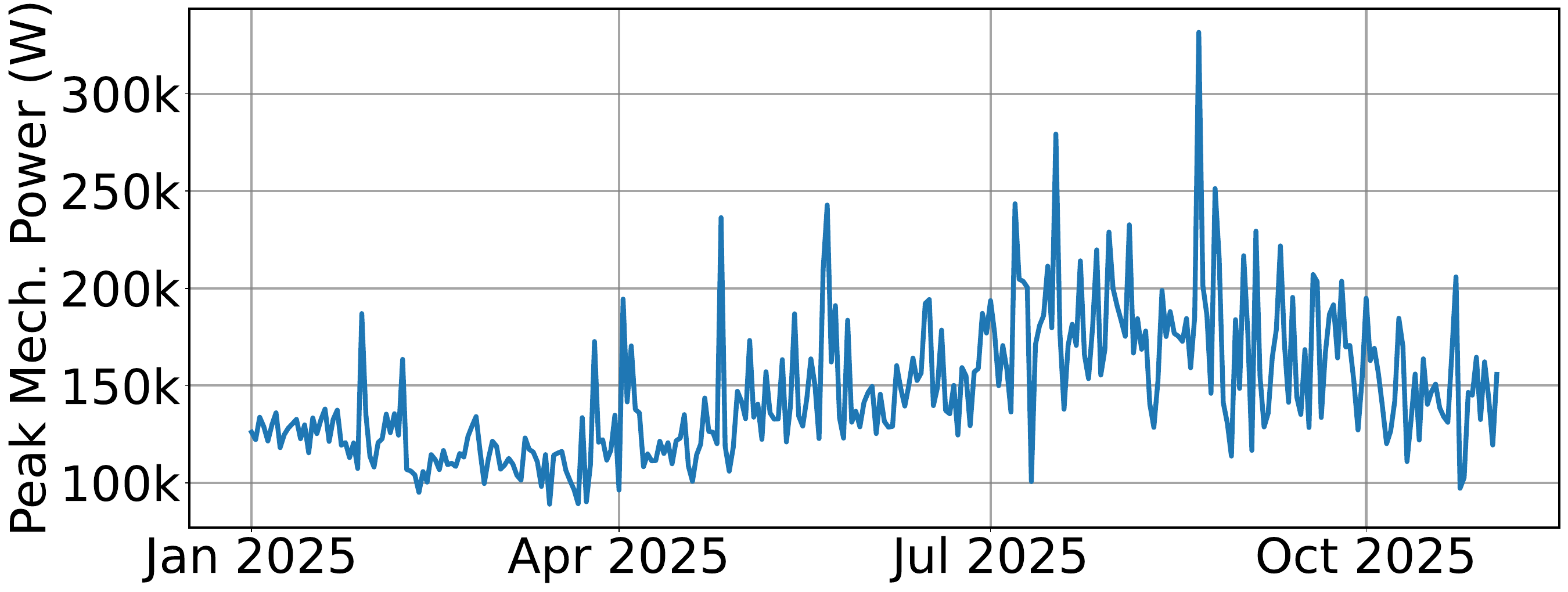}
    \caption{Daily mechanical peak-minute power for the first ten months of 2025 in one MSB.} 
    % Hotter summer days demand more power for cooling, and for some minutes the power can breach the 300~kW estimated for planning.}
    \label{fig:hmd_power_plot}
    \vspace{-3mm}
\end{figure}

\noindent \textbf{Seasonal and Peak Loads.}
Figure~\ref{fig:hmd_power_plot} shows the daily peak-minute mechanical power usage for one MSB over several months. The data illustrates that during north hemisphere summer, cooling demand increases, and for short intervals, the power required for mechanical systems can exceed the levels estimated during initial planning. This variability must be accounted for in the design and operation of the datacenter to ensure that sufficient power is always available for both IT and cooling, even under peak load conditions.

\subsection{The Three Phases}

\vspace{2pt}

\noindent \textbf{Provisioning Phase.}
Before new AI accelerators are deployed at scale, the datacenter must undergo a power provisioning phase. 
During this stage, infrastructure teams use projection models to estimate the power and performance characteristics of the upcoming hardware. 
The main goal is to determine how many accelerators can be supported within the available power budget. 
This involves setting an initial per-accelerator power allocation and defining the operational power-limit range for the accelerators. These early decisions shape the cluster’s architecture and set the power-performance upper bound. \textbf{Deployment Validation Phase.} Once the hardware is deployed, we have the opportunity to validate and refine the assumptions established during provisioning. Datacenter operators can analyze power consumption data and workload performance to make informed adjustments to GPU power limits and margins across power delivery hierarchy, ensuring they align with real-world conditions. \textbf{Operational Phase.} In the operational phase, the focus shifts to real-time power management with the objective of capturing the last percentage of stranded power, while ensuring safety of power delivery system.

\section{Phase 1: Planning and Provisioning}
\label{sec:power_provision}

\vspace{2pt}

\subsection{Model and Methodology}
The primary goal during the power provisioning phase is to determine the optimal cluster configuration that maximizes overall throughput ($T(p)$) at the datacenter level within the fixed power budget ($\Ptot$). This process requires selecting the total number of accelerators, which directly determines the permissible power limit ($p$) for each unit and the resulting power allocated per rack. Inputs to this optimization process include the operational power-limit range of the accelerators, their performance characteristics ($f(p)$) at different power levels, 
and both rack-level power consumption and cluster-wide performance. Note that at this stage the accelerator is not operationalized and these inputs are mainly based on predictive models.
Overall, we formulate the goal as: 

\begin{equation} 
\max_{p} \quad T(p) = N(p) \cdot f(p) 
\end{equation} 

where $N(p) = \left\lfloor \dfrac{P_{\text{total}}}{g(p)} \right\rfloor$ is the number of GPUs that fit in the power budget.

\begin{equation}\label{eq:g}
    g(p) \;=\; \frac{1}{\delta}\!\left(\, p \;+\; \frac{P_{\mathrm{fix}}}{n_r}
          \;+\; \Pnet \,\right),
\end{equation}

where $\delta \approx 0.90$ is a de-rating factor (simultaneous peak across all components is unlikely). Note that $g(p)$ is affine in $p$ (hence convex). The number of GPUs that fit within the power budget is:
\begin{equation}\label{eq:N}
    N(p) \;=\; \min\!\left(\left\lfloor \frac{\Ptot}{g(p)} \right\rfloor,\;
               N_{\max}\right).
\end{equation}

This reduces to \textbf{maximizing performance-per-watt} $\eta(p) = f(p)/g(p)$, which is quasiconcave, meaning it has a single peak.

\begin{equation}\label{eq:relaxed}
    \max_{p \in [\pmin, \pmax]} \quad T(p) \;=\; \Ptot \cdot \frac{f(p)}{g(p)}
    \;=\; \Ptot \cdot \eta(p).
\end{equation}
%

%Concavity arises because: (i)~achieved GPU flops can scale sublinearly with
%power; (ii)~HBM bandwidth is constant above ${\sim}1000$\,W; and (iii)~communication-bound phases are power-insensitive.

Additionally, the local constraints due to power delivery hierarchy also need to be modeled according to Equation~\ref{eq:opt}.

\begin{equation}\label{eq:opt}
    \max_{\{p_k\}}  \sum_{m \in \mathcal{M}} \;\sum_{s \in \mathcal{S}_m}\;\sum_{r \in \mathcal{R}_s} \;\sum_{k \in \mathcal{K}_r} n_k \cdot f(p_k) 
\end{equation}

Equation~\ref{eq:opt} is solved subject to the following constraints:
\begin{gather*}
    \sum_{k \in \mathcal{K}_r} q_k(p_k) \;\leq\; C_r \forall\, r \in \mathcal{R}_s,\; s \in \mathcal{S}_m,\; m \in \mathcal{M} \text{(RPP)} \\
    \sum_{r \in \mathcal{R}_s} \sum_{k \in \mathcal{K}_r} q_k(p_k) \;\leq\; C_s  \forall\, s \in \mathcal{S}_m,\; m \in \mathcal{M} \text{(SB)} \\
    h_m + \sum_{s \in \mathcal{S}_m} \sum_{r \in \mathcal{R}_s}
        \sum_{k \in \mathcal{K}_r} q_k(p_k) \;\leq\; C_m
        \forall\, m \in \mathcal{M}
        \text{(MSB)} \\
    \pmin \;\leq\; p_k \;\leq\; \pmax
         \forall\, k
         \text{(GPU bounds)}
\end{gather*}

\begin{table}[h]
\caption{Definition of Notation}
\centering
\resizebox{\columnwidth}{!}{%
\begin{tabular}{@{}cl@{}}
\toprule
\textbf{Symbol} & \textbf{Definition} \\
\midrule
$p$               & Per-GPU power limit (decision variable), $p \in [\pmin, \pmax]$ \\
$f(p)$            & End-to-end per-GPU performance at power limit $p$ \\
$g(p)$            & Total datacenter power consumed per GPU (including overhead) \\
$N(p)$            & Number of GPUs deployable within $\Ptot$ \\
$T(p)$            & Aggregate cluster throughput \\
$\eta(p)$         & Performance-per-watt ratio \\
$\Ptot$           & Total datacenter IT power budget (constant) \\
$\alpha$          & GPU fraction of total rack power \\
$\delta$          & De-rating factor ($\approx 0.90$) \\
$\Pnet$           & Per-GPU network power allocation (constant) \\
$P_{\mathrm{fix}}$& Fixed per-rack non-GPU power (CPUs, fans, NICs) \\
$n_r$             & Number of GPUs per rack \\
$N_{\max}$        & Physical upper bound on GPU count (switch radix, floor space) \\
$\mathcal{M}$    & Set of MSBs (Main Switchboards) \\
$\mathcal{S}_m$  & Set of SBs (Switchboards) under MSB $m$ \\
$\mathcal{R}_s$  & Set of RPPs (Reactor Power Panels) under SB $s$ \\
$\mathcal{K}_r$  & Set of racks under RPP $r$ \\
$C_m,\, C_s,\, C_r$ & Power capacity of MSB $m$, SB $s$, RPP $r$ \\
$h_m(t)$         & Mechanical/cooling power for MSB $m$ (time-varying) \\
$p_k$            & GPU power limit for rack $k$ (decision variable) \\
$n_k$            & Number of GPUs in rack $k$ (fixed by rack type) \\
$q_k(p_k)$       & Total power draw of rack $k$ at GPU power limit $p_k$ \\
\bottomrule
\end{tabular}
}
\end{table}
By considering the empirical data (Section~\ref{sec:edata}), we arrive at p=80\% power limit to be the power limit that maximized cluster throughput. Concavity arises because: (i)~achieved GPU flops scales sublinearly with power; (ii)~HBM bandwidth is constant above ${\sim}1000$\,W; and (iii)~communication-bound phases are power-insensitive. While the exact curves are accelerator-dependent, the methodology is generalizable. We have applied the same approach to AMD GPUs and our internal AI accelerator.

\subsection{Empirical Data}
\label{sec:edata}
\begin{figure}[!htb]
\vspace{-3mm}
    \centering
    \includegraphics[width=\linewidth]{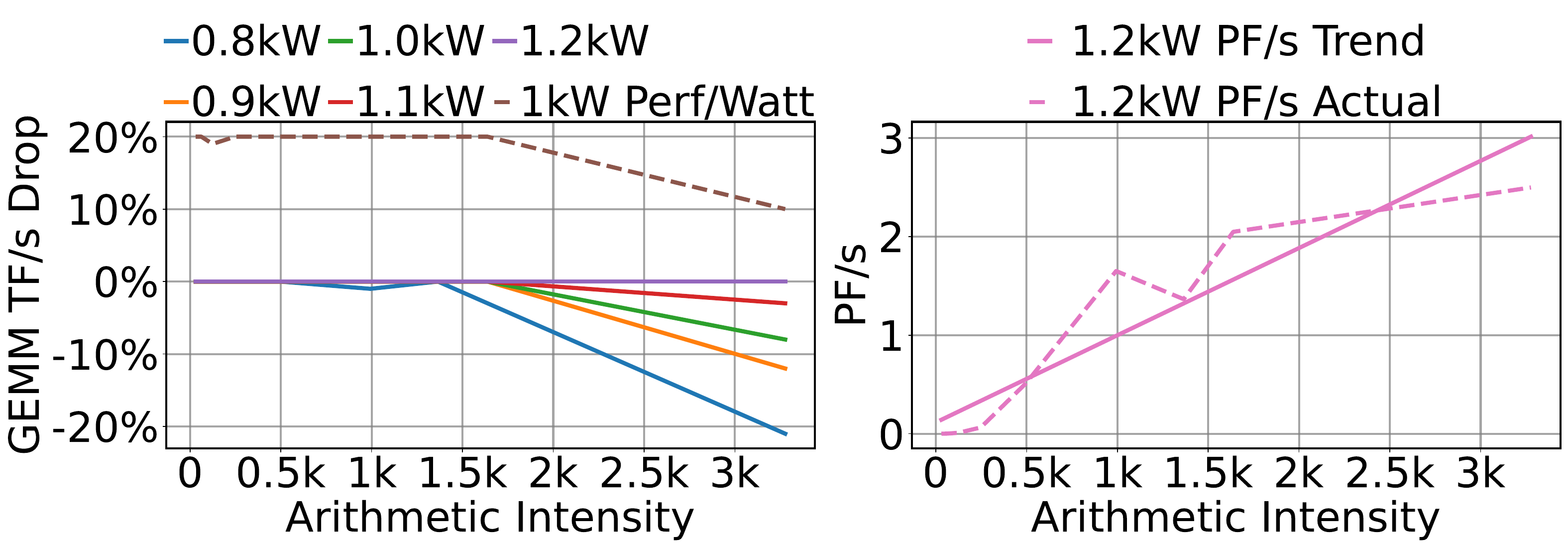}
    \caption{GB200 FP8 FLOPS sensitivity to power limit. Arithmetic Intensity = FLOPS/(Operands Bytes)}
    \label{fig:gemm_tdp}
\vspace{-3mm}
\end{figure}

\begin{figure}[!htb]
\vspace{-3mm}
    \centering
    \includegraphics[width=0.9\linewidth]{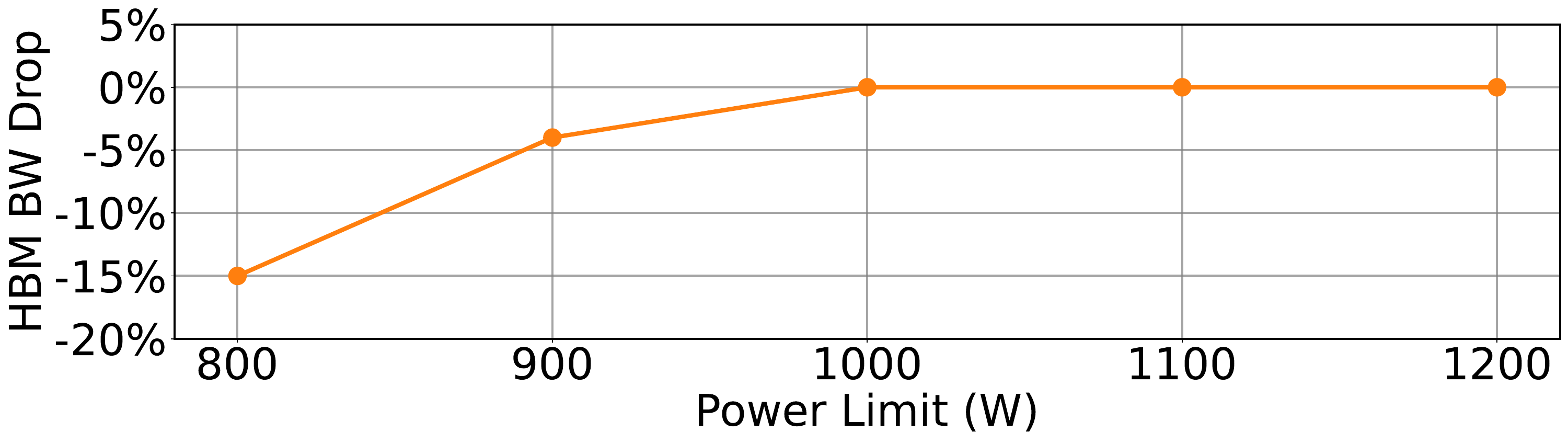}
    \caption{GB200 HBM bandwidth sensitivity to power limit.}
    %\todo{Update figures}}
    \label{fig:bw_tdp}
    \vspace{-3mm}
\end{figure}

\noindent \textbf{Density and Performance Trade-off.}
A central challenge in this process is balancing the accelerator density against individual performance. Within a fixed power envelope, increasing the number of accelerators means that each must operate at a lower power limit. As shown in Figure~\ref{fig:gemm_tdp} and Figure~\ref{fig:bw_tdp}, reducing power leads to lower sustained compute clock frequencies and, consequently, reduced computational throughput (FLOPS) and memory bandwidth for each GPU. 

Hence, the task is to find the configuration in which the aggregate throughput from additional accelerators outweighs the loss in per-accelerator performance. 
The optimization objective is to maximize total cluster performance: the product of the number of accelerators and the per-accelerator performance at a given power level. The constraint is that the total power consumed by all accelerators at a given power hierarchy level remains within the power budget for that level.

% The optimization objective is to maximize $Perf_{Cluster}$, subject to the constraint that the total power consumed by all accelerators and supporting systems does not exceed the available data center power budget ($P_{budget}$):

%\begin{equation}
%\label{eq:cluster_perf}
%    Perf_{Cluster} = num_{acc} \times perf_{acc}(power\_limit)
%\end{equation}

\begin{comment}
\vspace{2pt}
\noindent \textbf{Modeling Rack Power Consumption.}
To accurately project rack power, we model total rack consumption as the sum of the power drawn by each component, weighted by a utilization factor specific to that component. These factors are derived from empirical workload analysis.
For instance, we assume 33\% utilization for CPUs and 60\% for SSDs. 
Additionally, we apply a global de-rating factor of 90\% to the total, reflecting the reality that not all components operate at peak power simultaneously in production.
Notably, GPUs account for more than 70\% of total rack power.
\end{comment}

\vspace{2pt}

\noindent \textbf{Performance Sensitivity to Power Limit.}
Figure~\ref{fig:gemm_tdp} depicts the sensitivity of FLOPS for different power settings, across a range of General Matrix Multiplication (GEMM) shapes with increasing arithmetic intensity. The computations are with FP8 precision, although the trend applies to all the tensor-core supported precisions. 
The data indicates lack of sensitivity to power for arithmetic intensity lower than $\approx 1500$. 
As shown in the trendline of absolute FLOPS for 1200 W in the figure, such computations are less likely to fully utilize the hardware and reach peak FLOPS. 
Hence, even for higher arithmetic intensity operation, the {\pw} remains positive. Figure~\ref{fig:bw_tdp} shows the sensitivity of HBM bandwidth to power limits.

% Overall, the data indicate with power $\approx 1000W$ to be the sweet spot for \pw~ with no impact to HBM bandwidth (Figure~\ref{fig:bw_tdp}). 

\begin{figure}[!htb]
\vspace{-2mm}
    \centering
    \includegraphics[width=0.9\columnwidth]{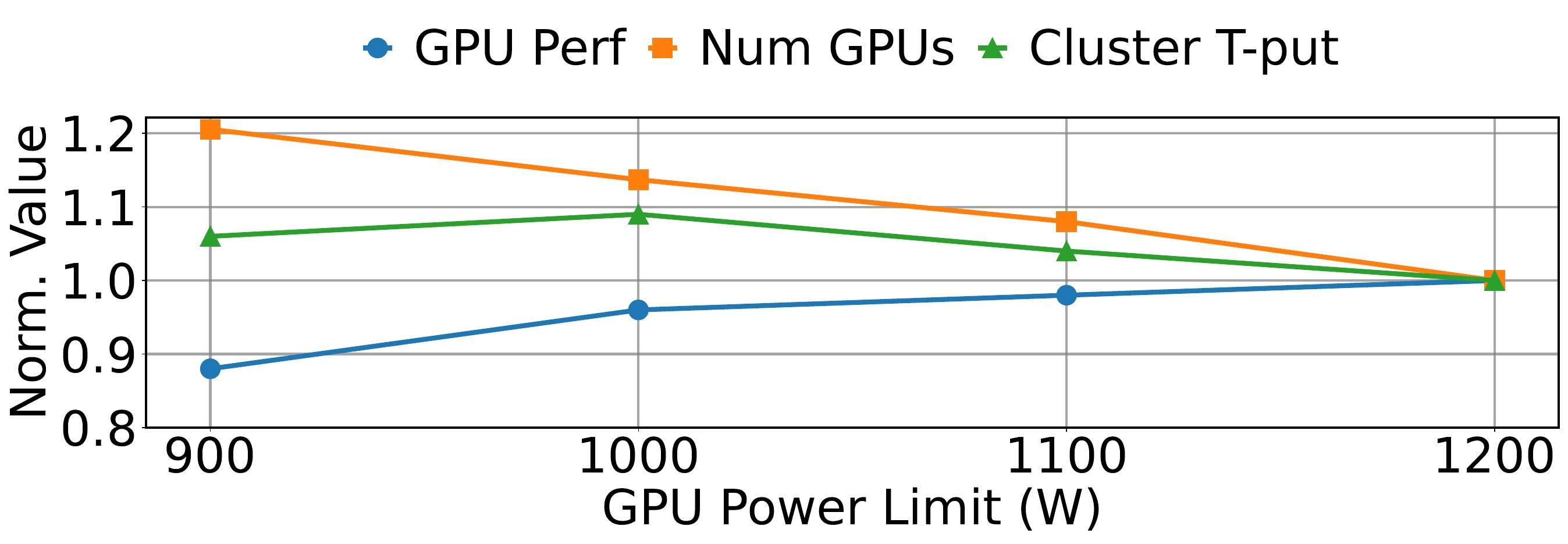}
    \caption{At a fixed power budget, higher GPU power increases per-GPU performance but not overall cluster throughput.}
    \label{fig:cluster_throughput_tdp}
\vspace{-2mm}
\end{figure}

We use performance projections based on simulations to analyze how changes in accelerator power affect computational throughput for large-scale pre-training workloads. This is done based on our internal graph execution simulation of the workloads, with different inputs for varying power limit.  Figure~\ref{fig:cluster_throughput_tdp} shows the impact of varying per-GPU power on three metrics: per-GPU performance, the number of GPUs that can be deployed within a fixed cluster power budget, and the resulting total cluster throughput. 
All values are normalized to the case where each GPU operates at the max 1200~W power.

We can see that the relationship between power and performance is non-linear at both the individual GPU and cluster levels. Reducing the per-GPU power from 1200~W to 1000~W results in only a 5\% decrease in per-GPU performance while reducing power draw by 16.7\%. Lowering the power to 900~W yields a 12\% drop in per-GPU performance, but saves 25\% in power. 
Importantly, operating at lower power allows more GPUs to be deployed within the same total power envelope. 

As a result, aggregate cluster throughput increases when the per-GPU power limit is set below the maximum, with commensurately higher number of GPUs landed in the cluster. 
Specifically, setting it to 900~W power yields a 6\% improvement compared to the 1200~W baseline, while 
1000~W power limit improves total cluster throughput by 9\%. We determined that the optimal \emph{\pw} operating point for the cluster is achieved when GPUs power limit is set to approximately 1000~W, at which point the power savings compensate the modest reduction in per-GPU computational performance allowing increase of GPU density in the datacenter.

\vspace{2pt}
\noindent \textbf{Modeling Rack Power Consumption.}
To estimate rack power consumption, we model the peak sustainable rack power as the sum of the power drawn by each major rack component, with each component’s power multiplied by a utilization factor that reflects typical workload behavior. These utilization factors are based on empirical measurements and projections.

After summing the component-level contributions, we apply a global de-rating factor ($\approx 90\%$) to the peak sustainable power. This de-rating factor accounts for the fact that, in real-world production environments, it is highly unlikely that all components will simultaneously reach their peak utilization. Table~\ref{tab:gb200_rack_power} presents the breakdown of power per component, and our derating assumptions. 
GPUs are the dominant contributor to the rack total power, typically accounting for more than 70\% of total power. Hence, the GPU power setting has the largest impact on the overall rack power consumption.

\begin{table}[ht]
\centering
\caption{Catalina-GB200 provisioned rack power. LiC and AiC are Liquid Cooled and Air Cooled components.We are unable to disclose the values of cells marked with an asterisk (*) due to confidentiality requirements.}
\label{tab:gb200_rack_power}
\footnotesize
\resizebox{\columnwidth}{!}{%
\begin{tabular}{|p{13mm}|p{6mm}|l|p{6mm}|p{6mm}|c|}
\hline
\textbf{Component} &\textbf{Power (W)} & \textbf{Count} & \textbf{Derate \%} & \textbf{Power (W)} & \textbf{Comments} \\
\hline
\multicolumn{6}{|c|}{Server-level Components} \\
\hline
 GPU+HBM & 960 & 36 & 100\% & 34560 & LiC  \\
 CPU+DIMM & 100 & 36 & 100\% & 3600 & LiC \\
 Back. NIC & \oops{66} & 36 & 90\% & \oops{2138} & LiC, 2x CX7+optics / GPU \\
 Front. NIC & \oops{18} & 36 & 80\% & \oops{518} & AiC, 1x 200G NIC / GPU \\
 SSD & 15 & 36 & 60\% & 324  & AiC, 1xE1.S / GPU \\
 Misc  & 50 & 18 & 50\% & 450 & AiC, BMC, Boot Drive, etc. \\
 \hline
 NVSwitch  & 580 & 18 & 90\% & 9396  & LiC, 2 dies / tray \\
 OSFPs & \oops{4} & 162 & 90\% & \oops{583} & LiC \\
 Fabric Mg. & \oops{100} & 9 & 50\% & \oops{450} & AiC \\
 \hline
 VR Loss & \multicolumn{3}{|c|}{15\%} & 2436 & AiC, Non CPU-GPU \\ 
Fans & \multicolumn{3}{|c|}{7\%} &289 & \\
\hline
\multicolumn{6}{|c|}{Rack-level Components} \\
\hline
 FE + optics & \oops{350} & 2 & 100\% & \oops{700} & AiC, 2x Wedge400 \\
 Mg. logic & \oops{100} & 1 & 100\% & \oops{100} & AiC \\
 \hline
 PSU loss & \multicolumn{3}{|c|}{4\%}  & 2239&   \\
 \hline
 \multicolumn{6}{|c|}{Total Power (W)} \\
 \hline
 \multicolumn{2}{|c|}{Total DC} & \multicolumn{3}{|c|}{52820} &  \\
 \multicolumn{2}{|c|}{Total AC} & \multicolumn{3}{|c|}{55966} & AC-DC conv. loss\\
 \multicolumn{2}{|c|}{\textbf{Provisioned Power}} & \multicolumn{3}{|c|}{\textbf{49200}} & \\
 \multicolumn{2}{|c|}{Max AC} & \multicolumn{3}{|c|}{72124} & Without derating\\
 \hline
 \multicolumn{2}{|c|}{AALC Power} & \multicolumn{2}{|c|}{3\%} & 1585 & \\
\hline
\end{tabular}
}
\end{table}

\vspace{2pt}
\noindent \textbf{Back End Network Power Allocation.} Beside the IT racks, a considerable portion of the power is allocated to the back end network gear. Table~\ref{tab:be_power} list the bottom up approach for estimating the backend network portion of the power considering the number of switches and the average power consumption per switch.

\begin{table}[htbp] 
\centering \caption{Back-End network Provisioned Power} 
\footnotesize
\label{tab:be_power} 
\resizebox{\columnwidth}{!}{%
\begin{tabular}{|p{15mm}|p{8mm}|p{6mm}|p{8mm}|p{25mm}|} 
\hline 
\textbf{Cluster level (per 2x GB200 Rack)} & \textbf{Rack Power (kW)} & \textbf{Count} & \textbf{Power (kW)} & \textbf{Comments} \\ 
\hline 
 RS & 1.88 & 3 & 5.6 & 1st layer switch \\ \hline 
 FS & 1.88 & 0.5 & 0.9 & 2nd layer switch \\ \hline 
 SS & 1.99 & 2.25 & 4.5 & 3nd layer switch \\ \hline 
 BE Power per 2xIT rack & & & 11.1 & 11\% of IT racks and 8-9\% of 150MW power  \\ 
 \hline 
 \end{tabular} 
 }
 \end{table}

\vspace{2pt}
\noindent \textbf{Finalized Provisioned Rack Power.}
Table~\ref{tab:tdp_scenarios} compares datacenter power allocation and resulting cluster performance for two accelerator generations (H100 and GB200) under different power settings. Based on these data, we consider GB200 power limit of 960~W (80\% of the TDP) as the \emph{\pw} optimized setting to maximize the cluster performance.

\begin{table}[h!]
%\vspace{-3mm}
\centering
\caption{Data center power allocation and cluster performance for different accelerator generations and power settings. H100 numbers provided for reference.} 
%\vspace{-1mm}
\label{tab:tdp_scenarios}
\footnotesize

\resizebox{\columnwidth}{!}{%
\begin{tabular}{|l|p{9mm}|p{9mm}|p{9mm}|}
\hline
\textbf{Category} & \textbf{H100 (700W)} & \textbf{GB200 (960W)} & \textbf{GB200 (1200W)} \\ \hline
\textbf{DC Power (kW)} & 150,000 & 150,000 & 150,000 \\ \hline
\textbf{IT oversubscription (KW)} & 0\% & 8\% & 8\% \\ \hline
Physical Accelerator per Rack (\#) & 16 & 36 & 36 \\ \hline
Rack Power Budget (kW) & 17.7 & 49.6 & 56.2 \\ \hline
Network Power  & 6\% & 8\% & 8\% \\ \hline
Region Turn-up Power (kW)  & 10,000 & 10,000 & 10,000 \\ \hline
Support Serv. Power \% Server Power & 10\% & 10\% & 10\% \\ \hline
AALC Power \% of Server Power & 0\% & 3\% & 3\% \\ \hline
Power for Racks + AALC +  Support Serv. & 142280 & 135800 & 135800 \\ \hline
Total Rack Power (kW) & 128052 & 118146 & 118146  \\ \hline
Physical GPUs (K) & 108 & 86 & 74 \\ \hline
Per GPU Performance norm. to H100  & 1 & 2.4 & 2.5 \\ \hline
Aggregate GPU Perf norm. to H100  & 1 & 1.9 & 1.7 \\ \hline
%Cluster Training Perf norm. to H100 & 1 & 2.0 & 1.9 \\ \hline
\end{tabular}
}

\end{table}

At 960~W power limit, each GB200 delivers 2.4$\times$ the per-GPU performance of H100, and the overall cluster achieves 1.9$\times$ the throughput, even though fewer physical GPUs can be deployed within the same power budget. Compared to 1200W GB200, this is around 11\% increase in the cluster throughput. Note that due to other constraints (e.g. back end network switch radix), the exact number of GB200 deployed where around 83K.
% This improvement is due to both architectural advances and higher per-GPU efficiency in the GB200.

\begin{comment}
Within the GB200 generation, the choice of power has a first order impact on cluster-level performance.  Operating GB200 GPUs at the maximum 1200~W power yields the highest per-GPU performance, but the increased power draw limits the total number of GPUs that can be deployed, resulting in lower aggregate GPU performance than in  960~W configuration. 
\end{comment}

\vspace{2pt}

\noindent \textbf{Summary.} The results discussed in this Section highlight two insights:
First, upgrading to a new accelerator generation can double cluster performance within the same power envelope.
Second, careful selection of the per-GPU power limit is important; operating at a slightly reduced power can yield better overall efficiency and throughput than simply maximizing per-GPU performance.

\section{Phase 2: Deployment Validation}

\begin{comment}
The operational phase of a large-scale AI cluster introduces new challenges and opportunities for power management that extend beyond initial provisioning.
While early planning relies on conservative estimates and static models, operation exposes discrepancies between projected and actual power consumption due to dynamic workload behavior, environmental variability, and infrastructure constraints. 
Effective power management at this stage requires accurate, fleet-wide telemetry and adaptive strategies to address both static and dynamic imbalances in power distribution. 
\end{comment}

During operation, large-scale AI clusters face power-management challenges that static provisioning cannot fully capture, due to dynamic workloads, environmental variation, and infrastructure constraints. Safe and efficient power use requires accurate fleet-wide telemetry and strategies that exploit both temporal and spatial smoothing effects.

\noindent \textbf{Temporal averaging.} 
Circuit breakers trip based on overdraw duration as well as magnitude, effectively enforcing limits on moving-average power rather than instantaneous spikes. For example, an RPP may tolerate a 10\% overdraw for ~17 minutes but trips in ~60 seconds at 40\%, whereas an MSB may trip in ~60 seconds at only 15\% overdraw~\cite{dcsafepoweroversubscription,dynamoFB}. Operational power tuning must respect these time scales.

\begin{comment}
Each power device in a data center uses a circuit breaker to prevent damage from excessive power draw. If the device’s power consumption exceeds the breaker's rated capacity, the breaker will trip and disconnect the device from the power supply. However, this does not happen instantly. For example, an RPP can handle a 10\% overdraw for about 17 minutes before tripping, but a 40\% overdraw will cause it to trip in just 60 seconds. Lower-level devices like RPPs can tolerate higher and longer overdraws compared to higher-level devices such as MSBs. An MSB will trip after 60 seconds if the overdraw reaches only 15\%~\cite{dcsafepoweroversubscription,dynamoFB}. Such temporal window provides an organic smoothing, resulting in sensitivity in a moving average power over a window of a few seconds, rather than instantaneous power. Tuning of power at operational stage needs to take such temporal averaging scale into account.
\end{comment}

\noindent \textbf{Spatial averaging.} Peak power across GPUs and racks is not perfectly aligned and show millisecond divergence, even for
synchronous pre-training workloads. This lack of simultaneity
creates spatial smoothing that allows per-rack power to be
provisioned closer to a lower percentile of the aggregate
distribution rather than the absolute peak.

We next describe how we validate rack-level measurements, analyze power headroom imbalances across the hierarchy, and design operational policies that maintain safety while maximizing performance.

\begin{comment}
The transition of a cluster into its operational phase presents an opportunity for the empirical validation of assumptions made during the antecedent provisioning stage. Discrepancies between empirically observed power consumption and the theoretical worst-case utilization models may necessitate in-situ adjustments to operational TDP of the accelerators.
A primary contributor to power stranding—the under utilization of provisioned electrical capacity—is the static imbalance of loads across the data center's power hierarchy. This imbalance is an inherent consequence of the heterogeneous placement of different rack types (with different provisioned power) throughout the facility, and Section~\ref{sec:imbalance} provides the details. However, prior to that, we establish the power measurement methodology.
\end{comment}

\subsection{Rack Power Measurement Validation}

To ensure safe and efficient operation of large-scale AI clusters, we need to monitor rack-level power consumption in real time. 
% This is especially important for updating and validating the per-GPU power-limit values that were initially set during the provisioning phase, as actual workload behavior and environmental conditions may differ from early projections. 
Ideally, power telemetry should be both accurate and readily available across the entire fleet, 
enabling continuous feedback and dynamic adjustment of power allocations.

\begin{figure}[t]
    \centering
    \includegraphics[width=0.48\textwidth]{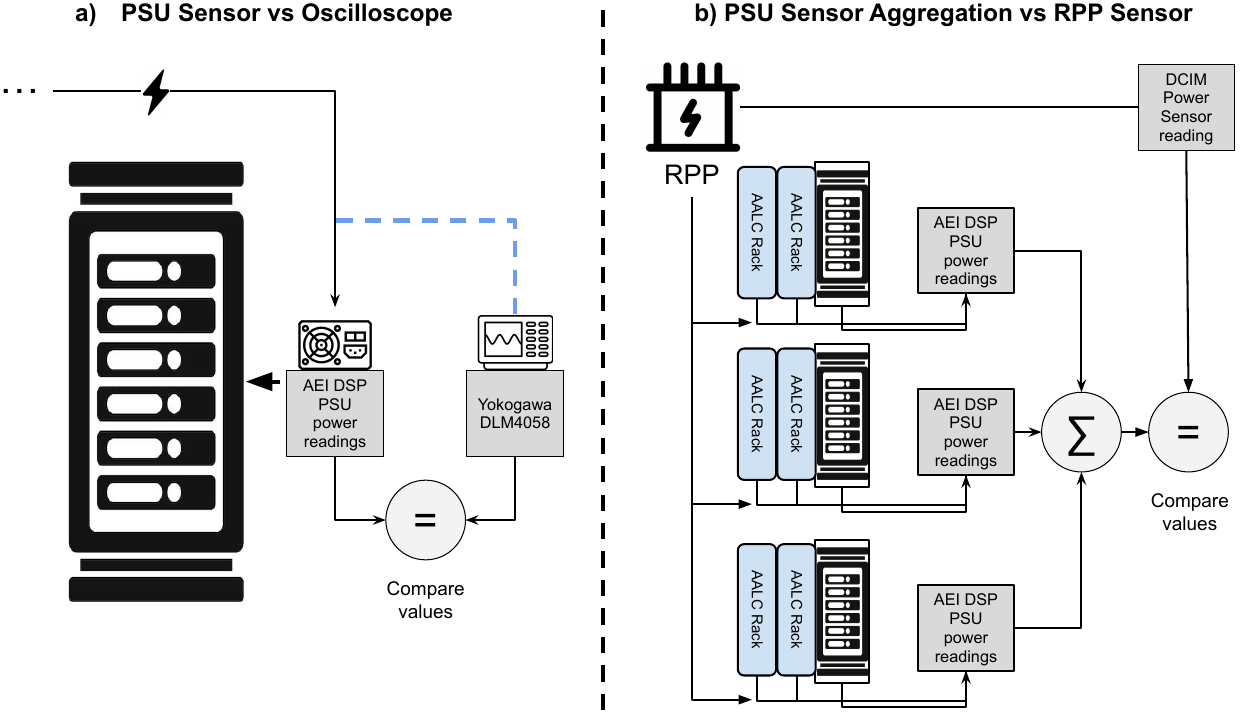}
    \caption{High-level diagram of PSU power validation. (a) Rack level: comparing a single rack PSU AC power sensor reading with oscilloscope. (b) RPP level: comparing RPP DCIM power sensor readings for RPPs using only IT racks and AALCs with aggregated rack PSU power sensors.}
    \label{fig:psu_validation}
\end{figure}

\noindent \textbf{Limitations of Readily Available Telemetry.}
% In practice, the most accessible source of rack power data is the AC power measurement reported by each rack’s power supply unit (PSU), which is collected and stored in Meta’s telemetry database.  
The primary source of rack-level power telemetry is the AC power measurement reported by each rack’s power supply unit (PSU). This measurement is obtained via a power metering integrated circuit (IC) embedded in the PSU, which samples voltage and current at regular intervals and computes the root mean square (RMS) power. 
The metering IC transmits these readings to a digital signal processor (DSP) every 100~ms, 
which applies running average over a one-second window to smooth out short-term fluctuations. 
The resulting averaged values are then sampled periodically (every few seconds) and logged for operational monitoring and analysis.
This process provides a continuous, fleet-wide view of rack power consumption.

However, we find that the PSU readings systematically overestimate true rack power consumption.
To quantify this bias, we compared PSU telemetry with two independent and more accurate sources: 
(1) high-frequency AC power measurements from an oscilloscope directly connected to the rack input, and 
(2) Data Center Infrastructure Management (DCIM) power sensors at the reactor power panel (RPP) level. 

\begin{figure}[h]
    \centering
    \includegraphics[width=0.95\linewidth]{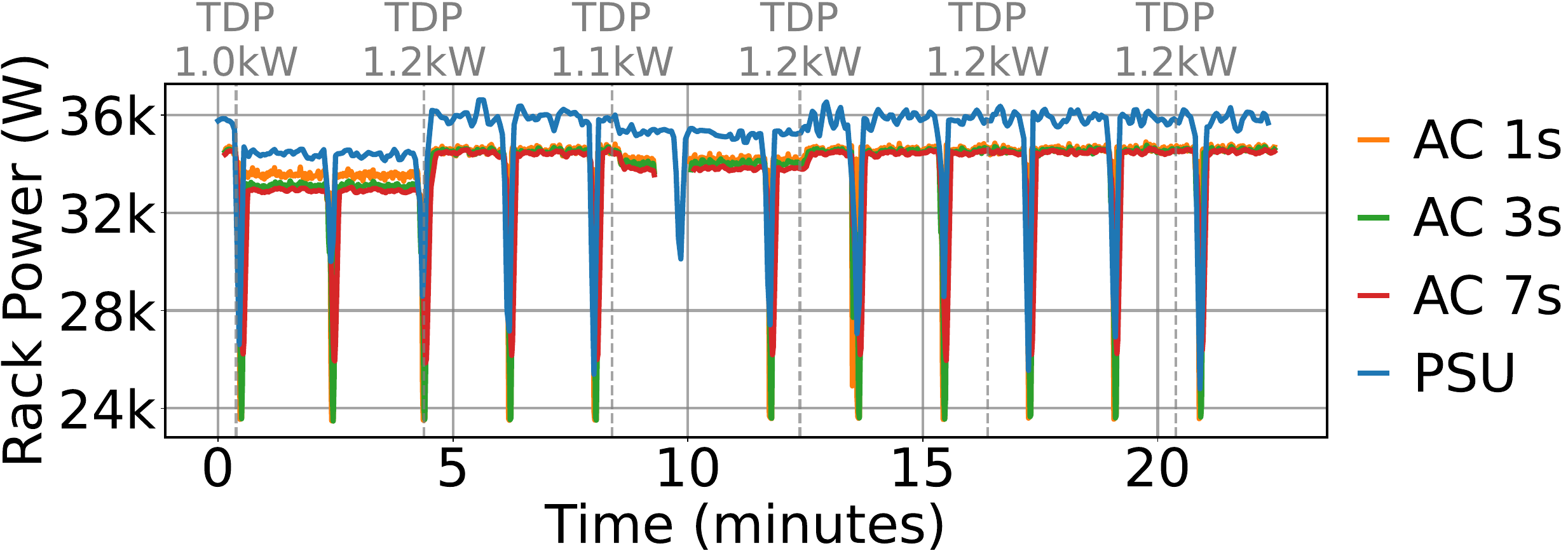}
    \caption{PSU versus AC oscilloscope power measurements.}
    \label{fig:psu_vs_oscilloscope}
\end{figure}

Figure~\ref{fig:psu_validation} presents a high-level diagram of the validation setup, while 
Figure~\ref{fig:psu_vs_oscilloscope} shows a direct comparison of PSU and oscilloscope power measurements. 
We can see that the PSU power values are consistently higher than both oscilloscope moving averages and DCIM sensor readings. 
This discrepancy is likely due to differences in averaging methods and the inherent conservatism of PSU telemetry, which is designed to avoid under-reporting that could lead to power overdraw.

\noindent \textbf{Correcting for Overestimation.}
While oscilloscope and DCIM measurements offer higher accuracy, the former is not scalable for fleet-wide monitoring due to its intrusive setup and limited deployment. In contrast, the latter lacks per-rack precision because it aggregates measurements from multiple racks. Therefore, we sought a method to adjust readily available PSU telemetry to better reflect actual rack power.

\begin{figure}[h]
    \centering
    \includegraphics[width=0.95\linewidth]{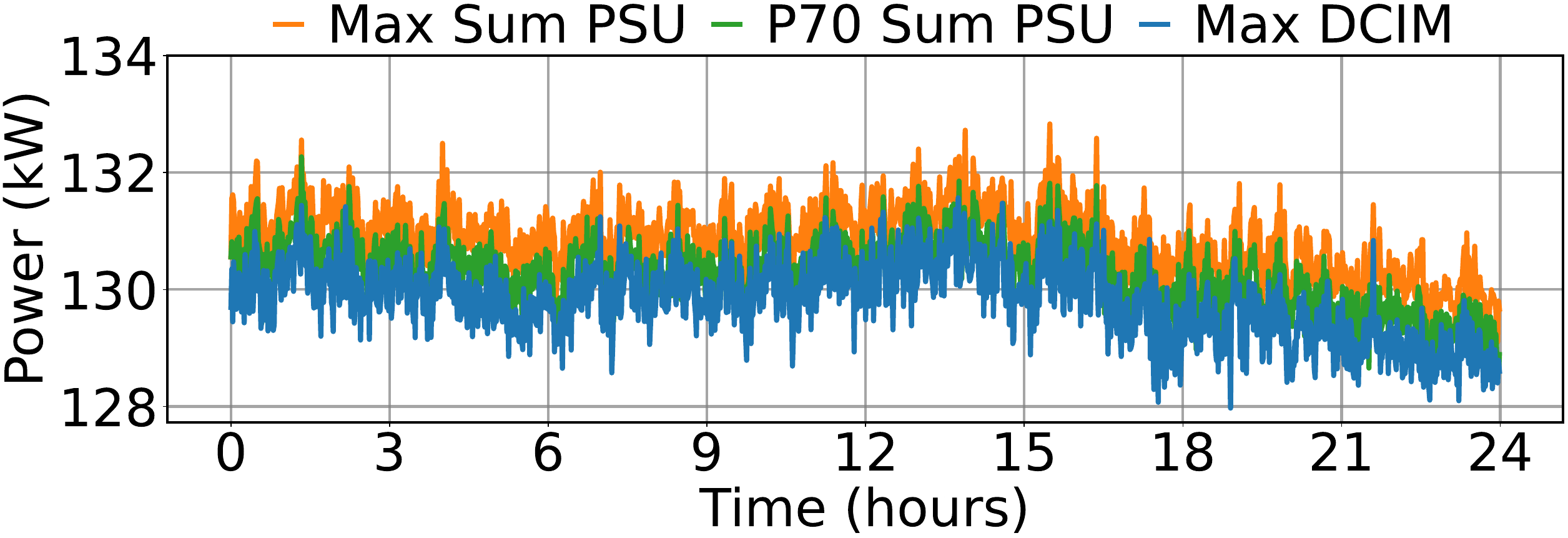}
    \caption{Comparison of maximum DCIM power samples with maximum and P70 of the sum of PSU power for one RPP.}
    \label{fig:dcim_vs_psu}
\end{figure}

\begin{figure}[h]
\vspace{-1mm}
    \centering
    \includegraphics[width=0.9\linewidth]{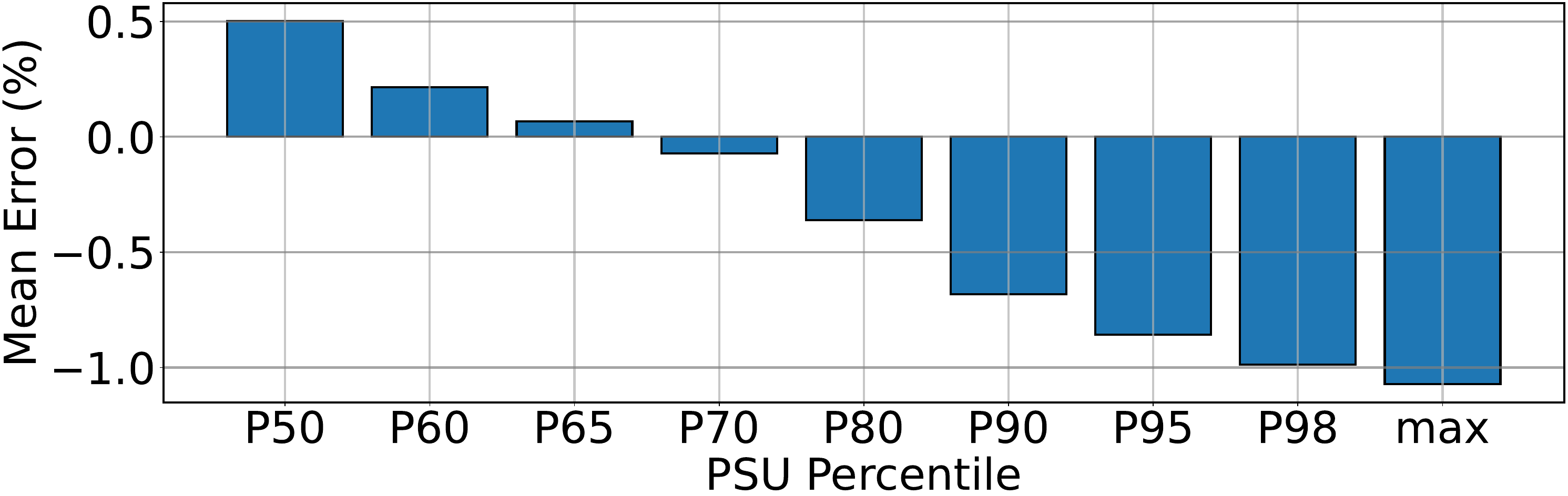}
    \caption{Comparison of maximum DCIM power samples with different aggregation statistics.}
    \label{fig:error_stats}
\end{figure}

By analyzing the distribution of PSU readings and comparing various aggregation statistics to DCIM data, 
we found that using the 70\textsuperscript{th} percentile (P70) of PSU power samples within each minute closely matches the DCIM sensor values and minimizes error. 
Figure~\ref{fig:dcim_vs_psu} shows the comparison of DCIM power samples with different PSU aggregation statistics for a representative RPP, and Figure~\ref{fig:error_stats} summarizes the mean error for each aggregation method. 
Adopting the P70 of PSU power as the standard aggregation method effectively mitigates the bias from using maximum PSU values, providing a more realistic and robust estimate of rack power while retaining the operational simplicity and scalability of PSU-based telemetry.

\noindent \textbf{Summary.}
While PSU power telemetry is convenient and widely available, it tends to overestimate true rack power. 
By adopting the P70 aggregation method, we achieve a practical balance between accuracy and scalability, enabling reliable rack power monitoring and dynamic TDP management.

\subsection{Static Power Imbalance in an AI Datacenter}
\label{sec:static_imbalance}
\noindent \textbf{Planned Power Headroom.}
Rack placement within a datacenter involves a complex set of constraints, including physical space, power connectivity, thermal limits, and network topology.
To assess the safety margin in power provisioning, we define the available \emph{planned power headroom} (PPH) as the difference between the rated power limit of a distribution device, 
such as an MSB or RPP,
and the sum of the provisioned power for all racks connected downstream.
%(Equation~\ref{eq:power_headroom}). 
%This headroom serves as a buffer to absorb transient power spikes and maintain operational reliability.
%
%\begin{equation}
%\label{eq:power_headroom}
%    PPH = \text{Limit} - \sum_{\text{racks under device}} (\text{provisioned power}_{\text{rack}})
%\end{equation}

\noindent \textbf{Sources of Imbalance.}
Two main factors drive the distribution of power headroom across the hierarchy: the final physical placement of racks and the distinct power budgets assigned to different rack types, including GPU compute racks, AALC racks, network racks, and supporting services racks. 
The interplay of these factors leads to considerable heterogeneity in available headroom at both the MSB and RPP levels.

\begin{figure}
\centering
\subfloat[Planned power headroom.]{
  \includegraphics[clip,width=0.48\columnwidth]{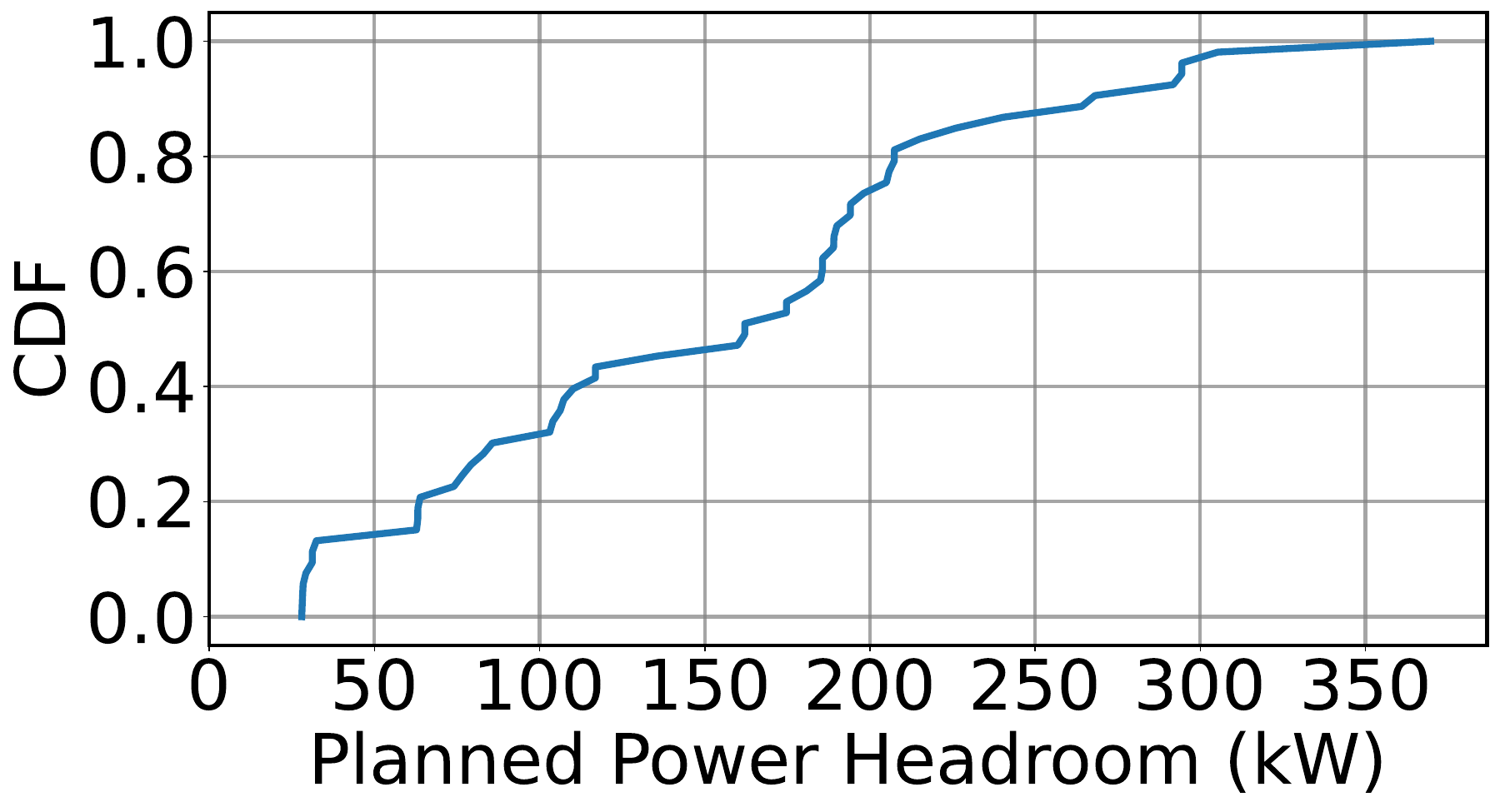}
  \label{fig:power_buffer_cdf_msb}
}
\subfloat[Planned power headroom per GPU.]
{
  \includegraphics[clip,width=0.48\columnwidth]{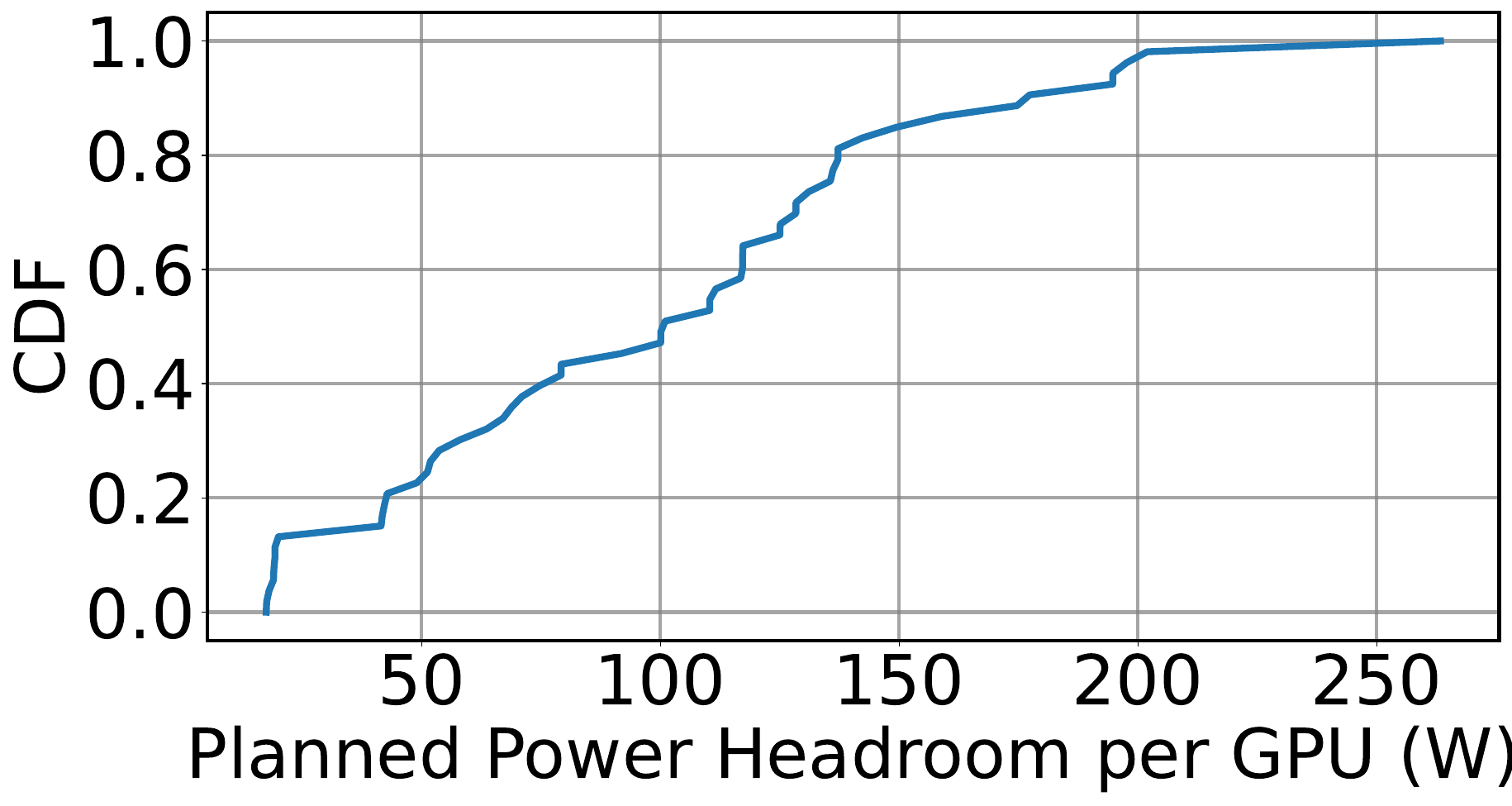}
  \label{fig:power_buffer_per_gpu_cdf_msb}
}
\caption{CDF of planned power headroom across MSBs: (a)~per-MSB level and (b) per-GPU level.}
\label{fig:cdf_msb}
\end{figure}

\begin{comment}
\begin{figure}
    \centering
    \includegraphics[width=1.0\linewidth]{figures/cdf_msb.png}
    \vspace{-6mm}
    \caption{Planned Power Headroom distribution across MSBs.}
    \label{fig:cdf_msb}
\end{figure}
\end{comment}

\begin{comment}
\begin{figure}
    \centering
    \includegraphics[width=1.0\linewidth]{figures/cdf_rpp.png}
    \vspace{-6mm}
    \caption{Planned Power Headroom distribution across RPPs.}
    \label{fig:cdf_rpp}
\end{figure}
\end{comment}

\begin{figure}
\centering
\subfloat[Power headroom per RPP.]{
  \includegraphics[clip,width=0.48\columnwidth]{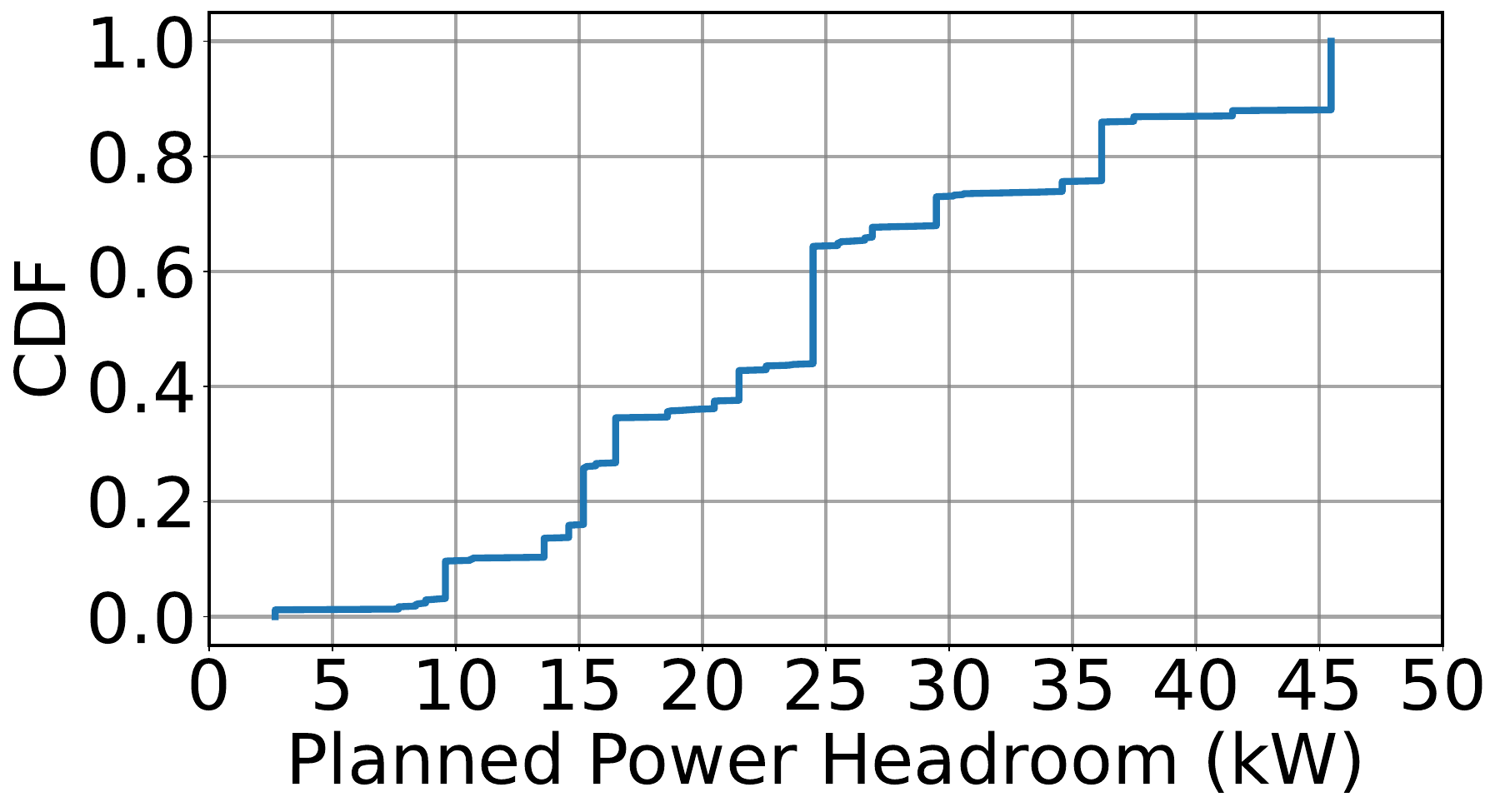}
  \label{fig:power_buffer_cdf_rpp}
}
\subfloat[Power headroom per GPU.]
{
  \includegraphics[clip,width=0.48\columnwidth]{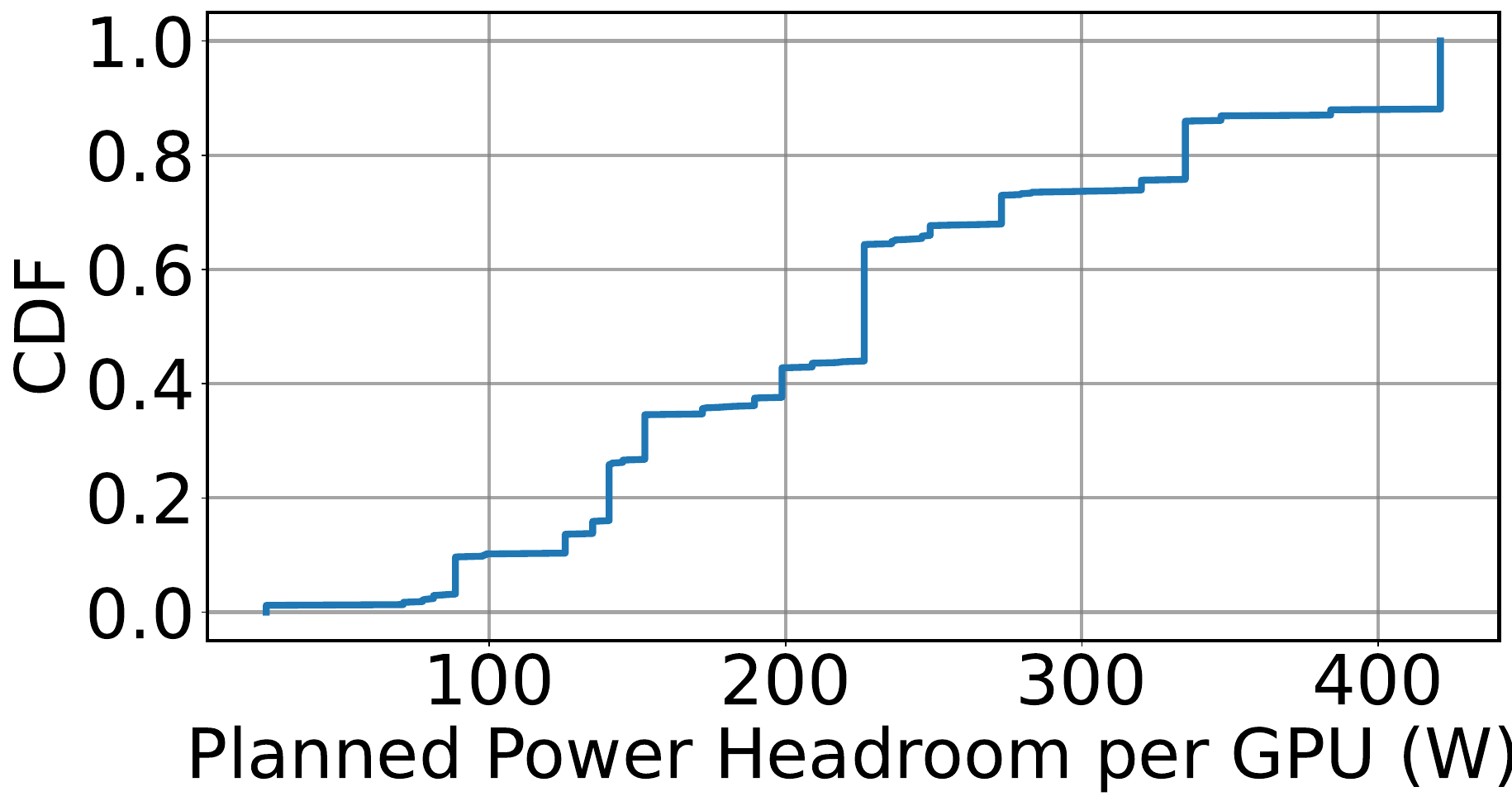}
  \label{fig:power_buffer_per_gpu_cdf_rpp}
}
\caption{CDF of planned power headroom across RPPs: (a) per-RPP level and (b) per-GPU level.}
\label{fig:cdf_rpp}
\end{figure}

\noindent \emph{Imbalance at the MSB Level.}
Figure~\ref{fig:cdf_msb} shows the cumulative distribution of planned power headroom across all MSBs assuming power budget of 3MW per MSB (Section~\ref{sec:powerhierarchy}).
In an ideal scenario with perfectly balanced rack placement and uniform power allocation, the CDF would appear as a sharp step function, with identical headroom across all MSBs. However, the actual distribution deviates from this ideal, revealing substantial heterogeneity. The curve shows that while some MSBs retain ample headroom, others operate much closer to their power limits. Specifically, while more than half of the MSBs maintain a headroom of at least 160~kW, a non-trivial fraction (13\%) operate with less than 50~kW of buffer. 
The mean headroom across all MSBs is $\approx$160~kW, which translates to only 100~W per GPU. This means static imbalance leave 5-10\% of power stranded. 
%This average, however, masks the underlying skew: several MSBs are highly constrained, and their limited headroom poses a risk of power bottlenecks if operational power is increased or if unexpected power excursion occur.

% \vspace{2pt}
% \noindent \textbf{Opportunities and Limitations for Mitigation.}

Some imbalance can be alleviated by relocating supporting services racks away from the most constrained MSBs, freeing up additional headroom. 
However, relocating AI compute racks is not practical at this stage, as it would require extensive fiber relaying and result in significant operational disruption. Consequently, the most constrained MSBs will continue to limit the cluster’s ability to safely increase GPU power limits or absorb transient power loads.

\noindent \emph{Imbalance at the RPP Level.}
Figure~\ref{fig:cdf_rpp} shows the corresponding headroom distribution at the RPP level, where each RPP has a power limit of 197.5~kW. The RPPs exhibit a more favorable profile, with an average headroom of 26~kW,
equivalent to over 200~W per GPU. 
This surplus indicates that, in contrast to the MSBs, RPPs are not the primary constraint in the power delivery hierarchy. Hence,
the MSB layer represents the main bottleneck for further scaling or aggressive power management strategies.
Some RPPs have virtually no margin for additional GPU capacity which can be mitigated by scheduling supporting services rack moves.

\noindent \textbf{Implications.}
Static imbalance could leave up to \textbf{10\% stranded power}. The observed power headroom heterogeneity shows that increasing GPU power or having aggressive performance targets must adhere to the most constrained MSBs, as they dictate the safe operation range for the entire deployment.

\subsection{Adjusting Operational Power Limit}
\begin{figure}
    \centering
    \includegraphics[width=1\linewidth]{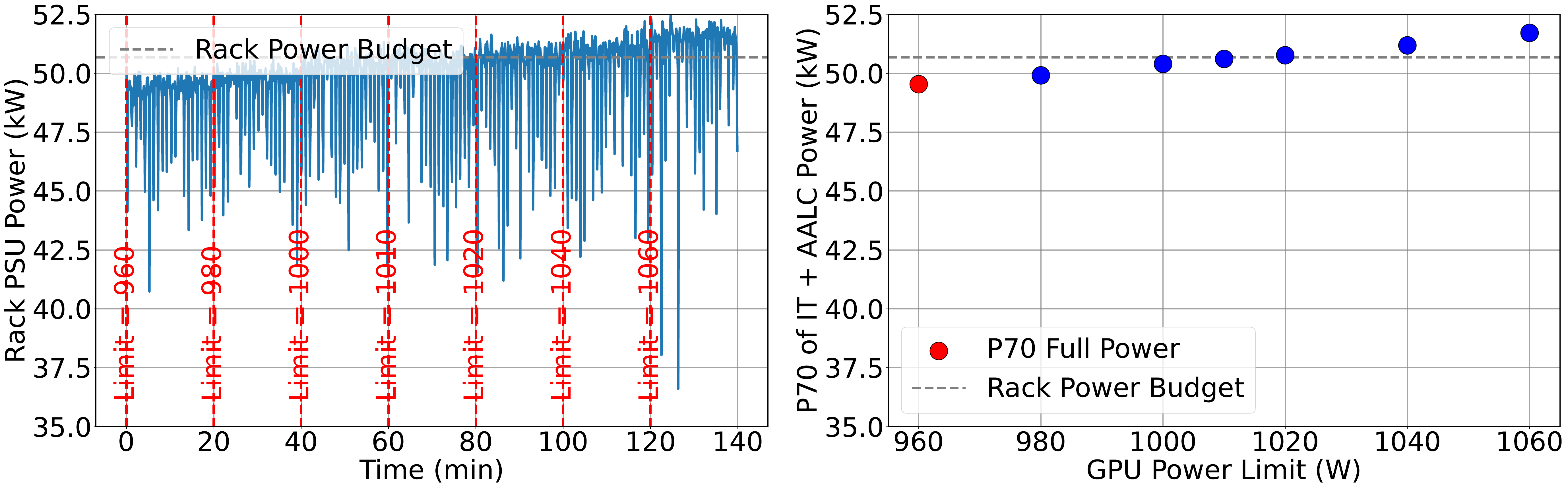}
    \caption{Power consumption of a GB200 rack running a dense Transformer model at a given GPU power limit.}
    % \cq{Change "TDP" in the right figure to "Power Limit"}}
    \label{fig:dense_model_power}
\end{figure}

Both static and dynamic power headroom, as well as the temporal and spatial scaling of power averaging, indicate the possibility of increasing the conservative 960~W per GPU established during initial provisioning. To confirm this possibility, we ran our benchmark model, and sampled power consumption every three seconds. The distribution and timelines are shown in Figure~\ref{fig:dense_model_power}.

We determined that a power limit of 1020 W ensures that the P70 closely matches the rack power budget. P70, as previously explained (see Figure~\ref{fig:error_stats}), minimizes the error between the sum of rack PSUs under a RPP and its DCIM power, while also correcting for the tendency of PSU power to be overstated. This increase in GB200 GPU power improves performance by $\approx 2.0-3.0\%$ on the expected workloads.

\subsection{Managing Power Swings}
The synchronous nature of contemporary distributed training algorithms, characterized by distinct periods of exposed communication that do not overlap with computation, induces large, synchronized power oscillations. When a considerable majority of GPUs in a cluster are allocated to a single pre-training job, these power fluctuations at the accelerator level aggregate, creating substantial power swings at the data center scale. The magnitude and frequency of these transients can pose a risk to the stability of the power utility grid and may potentially damage the power delivery infrastructure. 

\begin{comment}
    
% To mitigate these power swings, two primary approaches can be leveraged:
We explore two approaches to mitigate power swings: 

\vspace{2pt}

\noindent\textbf{A hardware-centric approach.} This utilizes features within the GB200 to enforce a configurable Minimum Power Floor (MPF). The raise MPF constrains the dynamic power range available for opportunistic boosting of compute-bound operators, which could result in a modest performance impact.

\vspace{2pt}

\end{comment}
We explore a software scheme to manage the power swings. This requires the workload to be aware of moments when power usage dips sharply before returning to higher levels. During these periods, the system can proactively launch dummy computational kernels to keep the GPU active. This ``gap filling" technique maintains a higher, more stable power draw, thereby smoothing the overall load profile. The focus of the solution is to reduce the transients up to a safe level and minimize the performance overhead. Two primary strategies were considered for implementing the power smoothing mechanism: a triggered, event-based approach (similar to \cite{power_stablizer}) and a continuously active, ``always-on" method. The event-based option was found to be excessively intrusive, as it would require substantial modifications to the core AI software stack, namely PyTorch. An alternative involving telemetry-based triggers was also evaluated, but the temporal resolution of our infrastructure's power monitoring was insufficient to initiate or terminate the smoothing process in a timely manner.
Given these implementation challenges and hardware limitations, we opt for the ``always-on" smoothing architecture.
% was selected as the most viable and non-intrusive solution.

\begin{figure}
    \centering
    \includegraphics[width=0.9\linewidth]{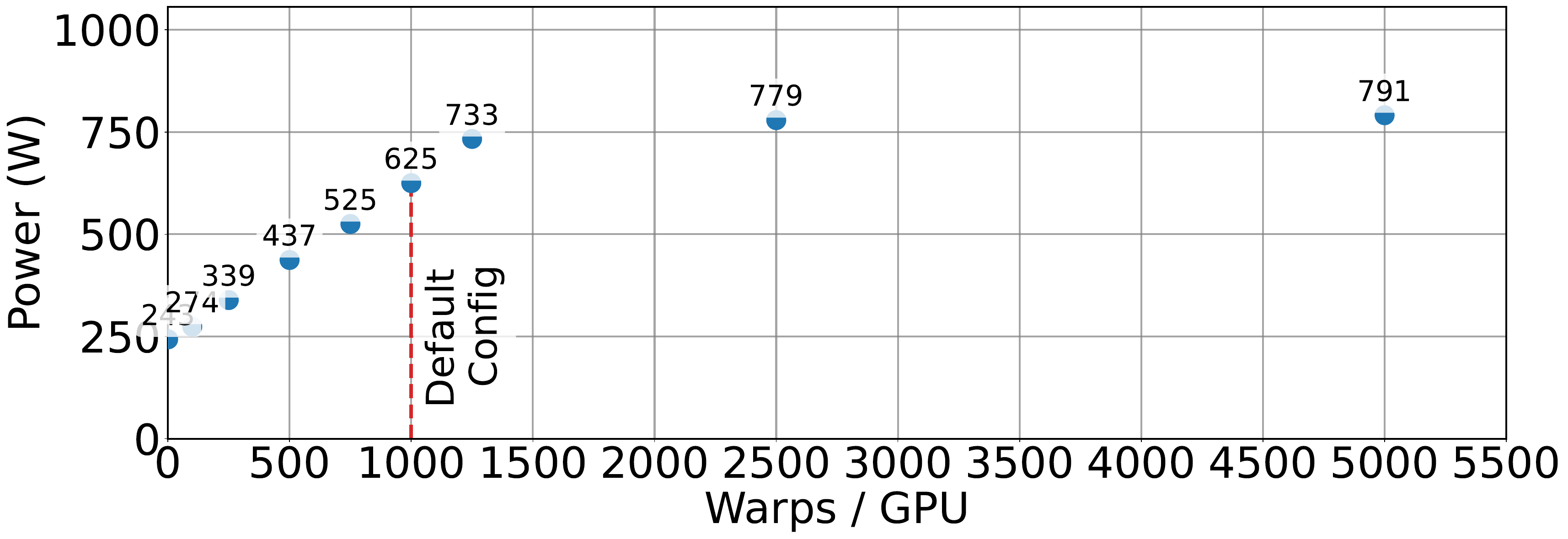}
    \caption{Software power smoother draws up to 800W of power.}
    \label{fig:ps_power}
    \includegraphics[width=0.85\linewidth]{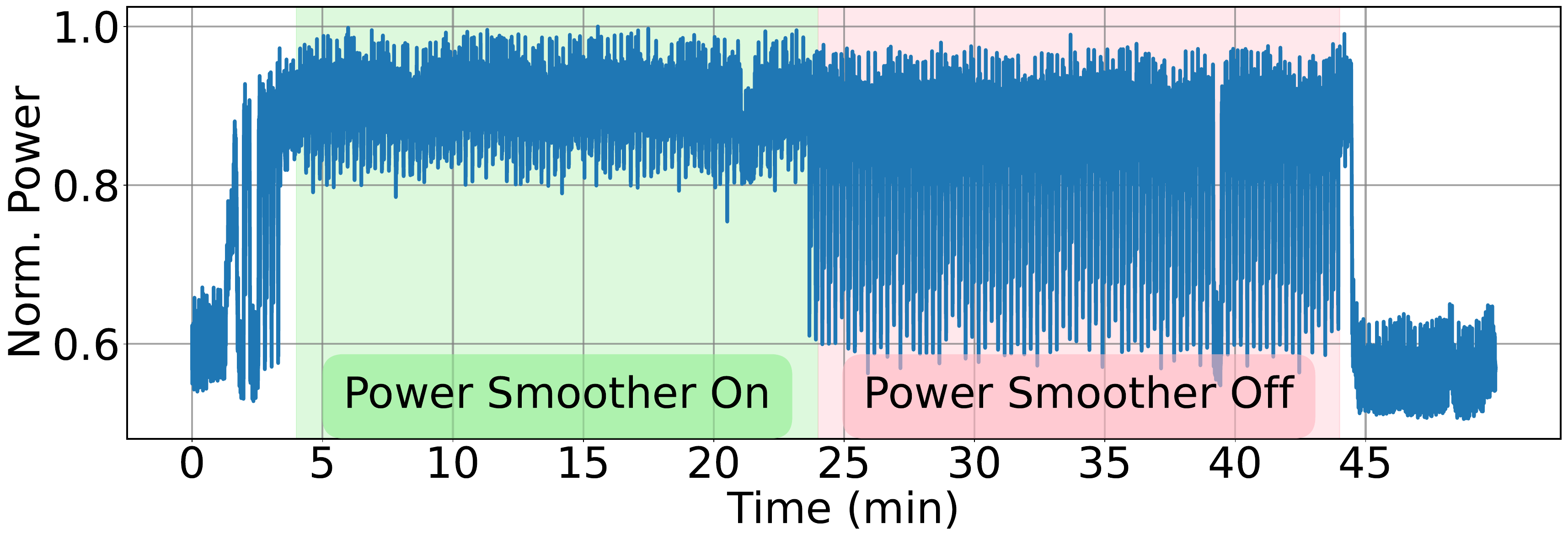}
    \caption{Software based power smoother on a region-scale training cluster. With the smoother, training related pulses are completely mitigated.}
    \label{fig:power_smoother}
\end{figure}

To minimize performance overhead on the primary training application, the power smoothing kernel is designed to be highly resource-frugal. It operates exclusively on data generated within registers, incurring no additional GPU memory footprint on L2 cache or HBM. The synthetic load is generated by issuing a continuous stream of instructions to a configurable number of warps per streaming multiprocessor, specifically targeting the Tensor Cores. As shown in Figure~\ref{fig:ps_power}, this method generates a load of nearly 800~W per GB200 accelerator.

To reduce overhead, the smoothing mechanism incorporates an adaptive backoff strategy. It continuously monitors the execution latency of its own Tensor Core instructions. If the latency surpasses a pre-calibrated threshold—indicating contention with the primary workload—the smoother automatically relinquishes control of that specific streaming multiprocessor. 
We empirically observed that this adaptive technique has less than 3\% impact on application performance while greatly simplifying software integration. 
Thus, the entire power smoothing feature is activated via a single environment variable, abstracting its complexity from the end-user.

\section{Phase 3: Active Operation Phase}\label{sec:dimmer}

\noindent\textbf{Why Dynamic Power Management.} As established in Section~\ref{sec:static_imbalance}, the observed power imbalance of up to 10\% across the delivery hierarchy represents a sizable source of stranded, underutilized power capacity. 
A dynamic power management framework enables a dynamic power limit strategy, and facilitates opportunistic performance boosting. Additionally, once such dynamic power managements are in place, we can consider a more aggressive number of accelerators in the DC in the datacenter planning phase.

\noindent\textbf{Dynamic Power Capping.}
Traditional power capping methods in datacenters are primarily implemented to maintain safe operation range, and not as a component to optimize power utilization. Such methods often respond to power limit violations by sharply dropping compute devices (e.g., CPU or GPU) power limit or forcibly evicting running tasks~\cite{dynamoFB,thunderbolt}. 
These abrupt actions can interrupt training jobs, waste compute resources, and undermine the reliability of large-scale AI workloads. 

The optimal strategy distributes power reductions evenly across all GPUs in a job. If we need to reclaim \textit{P} watts from a job spanning \textit{N} GPUs, reducing each GPU by \textit{P/N} is preferable to reducing only a subset \textit{$Q < N$} by \textit{P/Q}. The latter creates a deeper straggler and inflicts a disproportionate performance penalty. In practice, multiple jobs at different scales share power devices with varying margins, making the optimization non-trivial. Unless a workload is memory- or communication-bound, some degree of power capping is expected during operation.

To elevate dynamic power capping from a safeguard to an optimizer, we developed \emph{Dimmer} for training clusters. It adopts a more nuanced approach: when the total power draw of a power device approaches 97\% of its limit, the system gradually reduces GPU power across all racks within the affected power device. This strategy ensures that jobs continue executing, albeit at a marginally reduced performance, thereby preventing outright failure or restarts.

\emph{Dimmer} continuously records the power consumption of each device every second. These measurements are then averaged over a seven-second interval. This averaging technique helps to smooth out transient power spikes, enabling the system to distinguish between brief surges and sustained overages. Consequently, \emph{Dimmer} initiates power throttling only when a genuine, sustained overage is detected.

\begin{figure}[b]
    \centering
    \includegraphics[width=0.92\linewidth]{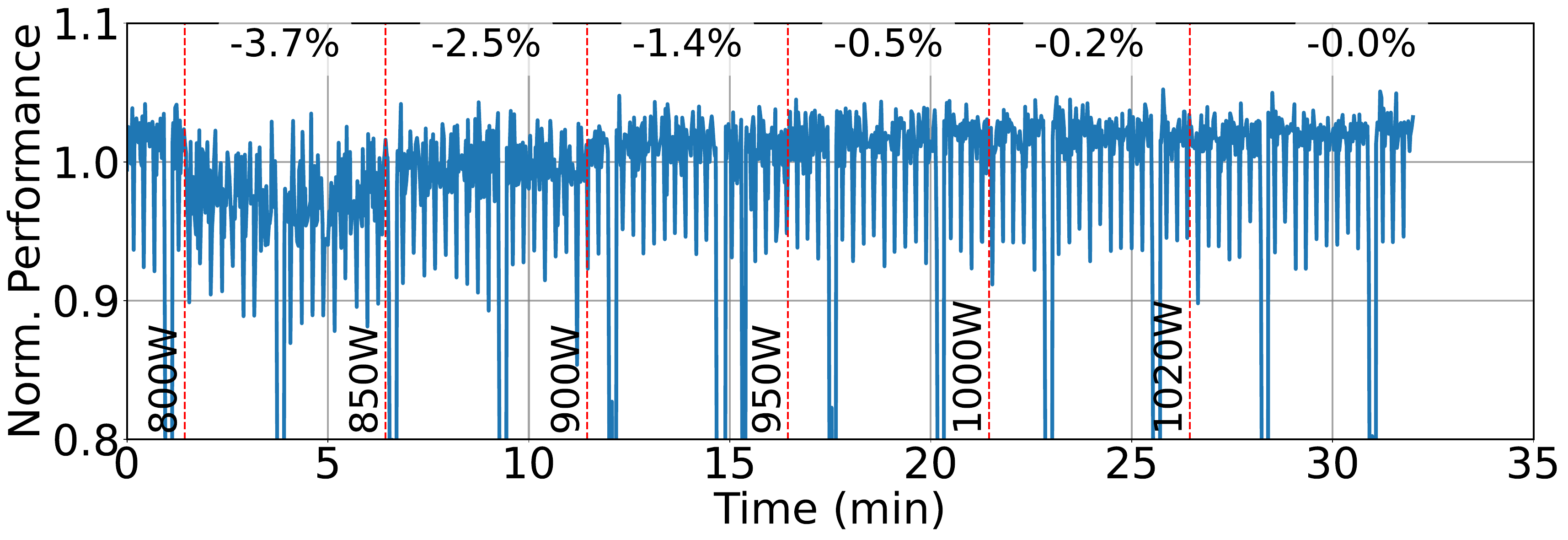}
    \caption{Adjusting power limit for a host in a training job, and the indirect impact on job performance and power draw. The y-axis is normalized power of the job excluding the host for which the power limits changes. }
    \label{fig:indirect_capping}
\end{figure}

The choice of a seven-second averaging period is motivated by the characteristics of circuit breakers, which protect each power device from power overages. Circuit breakers do not trip immediately upon exceeding the threshold\cite{dynamoFB,dcsafepoweroversubscription}. For instance, MSBs can tolerate power spikes up to 1.2$\times$ their rated limit for approximately 45 seconds, or up to twice the limit for about 30 seconds, before tripping. This inherent delay provides a window of time that allows the system to accommodate short-term spikes. Based on empirical testing, we found that a seven-second averaging window offers an effective and safe control period for managing power without causing unnecessary throttling. Algorithm~\ref{fig:indirect_capping} presents \emph{Dimmer}.

\begin{algorithm}
\caption{Dimmer Algorithm for a Power Device}

\begin{algorithmic}[1]
\scriptsize
\WHILE{system is running}
    \IF{currentDvcPwr $>$ limitDvcPwr}
        \STATE pwrToReclaim $\gets$ currentDvcPwr $-$ limitDvcPwr
        \FOR{pri \textbf{in} sorted(priorities)}
            \STATE ps $\gets$ $\sum_{s \in group}\text{avgPwr}_s$ 
            \STATE pls $\gets$ $\max($(ps - pwrToReclaim)$/$servers$_{pri}, 0)$
            \STATE r $\gets$pls / acceleratorCount
            \STATE dimmedTdp $\gets$ $\left\lfloor \frac{r}{10} \right\rfloor \times 10 + minTdp$
            \STATE pwrReduction $\gets$ numGPUs $\times$ (dimmedTdp $-$ minTdp)
            \FOR{s \textbf{in} group}
                \STATE e $\gets$ serverMinPwrCapped $+$ pwrReduction
                \STATE \texttt{dec}(pwrToReclaim, $\max(0, \text{avgPwr}_s -$ e$)$)
                \STATE serverCapList.append([sId, dimmedTdp])
            \ENDFOR
            \STATE capTime $\gets$ currentTime
            \IF{pwrToReclaim $\leq$ 0}
                \STATE \textbf{break}
            \ENDIF
        \ENDFOR
        \STATE PwrCapHosts(serverCapList)
    \ELSIF{capTime $+$ capExpirationTime $<$ currentTime}
        \STATE capTime $\gets \infty$
        \STATE uncapHosts(serversInPriority)
    \ENDIF
    \STATE sleep(decisionInterval)
\ENDWHILE
\end{algorithmic}

\label{alg_dimmer}
\end{algorithm}

% The y-axis is normalized power of the job excluding the host for which the power limits changes.

\noindent\textbf{Optimizing Power Capping for Distributed Training Jobs.}
In synchronous training, the overall job performance is determined by the slowest worker.
% , due to the tightly coupled nature of the computation. 
Hence, a power-capping event that throttles even a single GPU can create a ``straggler", which degrades the throughput of the entire job. This triggers a cascading effect: the performance, and thus the power consumption, of all other GPUs in the job is indirectly reduced as they wait for the throttled worker to finish. Figure~\ref{fig:indirect_capping} illustrates this scenario.

To mitigate this systemic inefficiency, the power management framework must be enhanced with job scheduler awareness. \emph{Dimmer} integrates scheduler metadata, such as the total size and resource distribution of jobs running under a power device subject to throttling, leading to more intelligent capping decisions. In Algorithm~\ref{fig:indirect_capping}, the job size is what defines the power capping priority. This scheduler-aware approach improves overall cluster throughput by, prioritizing the uninterrupted execution of large, mission-critical pre-training jobs, ensuring they are not disproportionately affected by a localized power-capping event on a single RPP.

\begin{figure}
    \centering
    \includegraphics[width=0.85\linewidth]{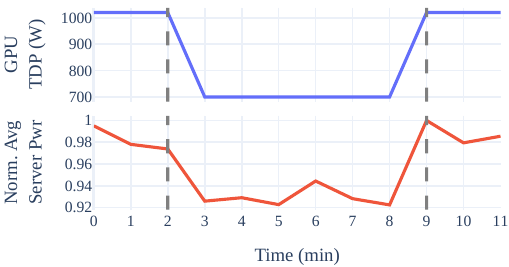}
    \caption{GPU TDP selected by Dimmer for servers running low-priority jobs and normalized average server power consumption for the servers running the low-priority jobs.}
    \vspace{-2mm}
    \label{fig:dimmer_case}
\end{figure}

\noindent\textbf{Dimmer Case Study.} We performed an experiment to illustrate the responsiveness of the Dimmer algorithm in reducing host power consumption under constrained conditions. Specifically, we artificially lowered the device power limit by 22\% compared to its original value illustrated in Figure~\ref{fig:dimmer_case}. At the two-minute mark, we simultaneously enforced this reduced power cap and initiated a high-power, high-priority workload on a subset of hosts belonging to a specific RPP. This high-priority load was sustained for one minute, while other jobs continued to run in parallel, sharing the same power device.

Figure~\ref{fig:dimmer_case} shows the GPU TDP selected by Dimmer for hosts running lower-priority jobs and the average power consumption of hosts with low power capping priority. 
The results show that Dimmer effectively responds to the power constraint by reducing the GPU TDP to the minimum allowed value, resulting in a roughly 7\% decrease in average host power of low priority jobs. Notably, although the high-power load lasts only one minute, the GPUs remain capped for an additional six minutes. 
Once the power cap timer expired, Dimmer restored the GPU TDPs to their normal level of 1020~W, indicating that the system successfully managed the temporary power surge and returned to standard operation after the cap period.

\noindent\textbf{Dimmer latencies.} \emph{Dimmer} retrieves power measurements through Nexu, a distributed polling system that collects energy and power data from electrical and mechanical devices using protocols such as Modbus~\cite{modbus}, BACnet~\cite{bacnet}, and OPC-UA~\cite{opcua}. Nexu uses a three-tier architecture: the \emph{Nexu Manager} assigns polling jobs to \emph{Nexu Workers}, which read devices at configurable intervals (as frequently as every 3 seconds for critical equipment) and push results to centralized stores, including the in-memory \emph{Nexu Aggregator} for low-latency access. This infrastructure enables real-time, scalable power monitoring that supports systems like Dynamo\cite{dynamoFB} and \emph{Dimmer}. %Figure~\ref{fig:nexu_latency} shows per-minute distributions of average and maximum read latency across devices.
The median average latency stays below one second, and the median maximum latency is slightly above one second, with rare outliers up to about 4.5 seconds. 
Read latency dominates the end-to-end control loop: the controller spends only a few milliseconds running the \emph{Dimmer} algorithm, and hosts need less than a second to receive commands and apply new GPU TDP limits.

\noindent\textbf{Reliability of Power management.} Transitioning the function of power capping from a static safeguard to a dynamic optimization tool carries two implications. First, this paradigm shift permits transient, localized power excursions that would be unsustainable at a global, cluster-wide level. This approach remains consistent with the initial provisioning strategy discussed in Section~\ref{sec:power_provision}. Second, power-capping events are transformed from emergency-driven occurrences into frequent operational adjustments. This necessitates robust fault tolerance. A failure in the power measurement instrumentation or the capping control logic must not expose the system to the unmitigated risk of tripping circuit breakers. To address this, a fail safe mechanism is implemented: each host system is designed to automatically revert to a predefined, globally safe power limit in the event that it no longer receives a periodic "heartbeat" signal from the dynamic power capping controller. This ensures system stability and prevents catastrophic failures even if the active control loop becomes unresponsive.

\begin{comment}
\lpiga{No-space to comment about this
\begin{itemize}
    \item BBU smart charging (https://fb.workplace.com/groups/385877898127498/permalink/27234042019551055/)
    \item Reliability aspects. what happens if Dynamo fails
\end{itemize}
}

\noindent\textbf{Summary.}
By enforcing power capping at the device level and applying adaptive, workload-aware policies, we ensure safe, reliable, and efficient operation of large AI training clusters. 
This approach minimizes job disruptions, maintains infrastructure stability, and provides clear performance expectations for service owners, even as workloads and environmental conditions fluctuate.
\end{comment}

\begin{comment}
    
\section{Dynamo Power Management Characterization}

\begin{itemize}
    \item Risks of exceeding power limits: fuse blowouts, shutdowns.
    \item Infrastructure protection mechanisms (e.g., Dimer systems).
    \begin{itemize}
        \item Describe the Dimmer Algorithm and the decisions it makes
    \end{itemize}
    \item Operational policies to prevent power violations.
    \item Brief overview of regional-level infrastructure (e.g., Dynamo X).
    \item How regional constraints and systems impact cluster deployment.
\end{itemize}
\end{comment}

\vspace{-3mm}
\section{Putting it all together}
\label{sec:results}
\begin{figure}[h]
    \centering
    \includegraphics[width=1.0\linewidth]{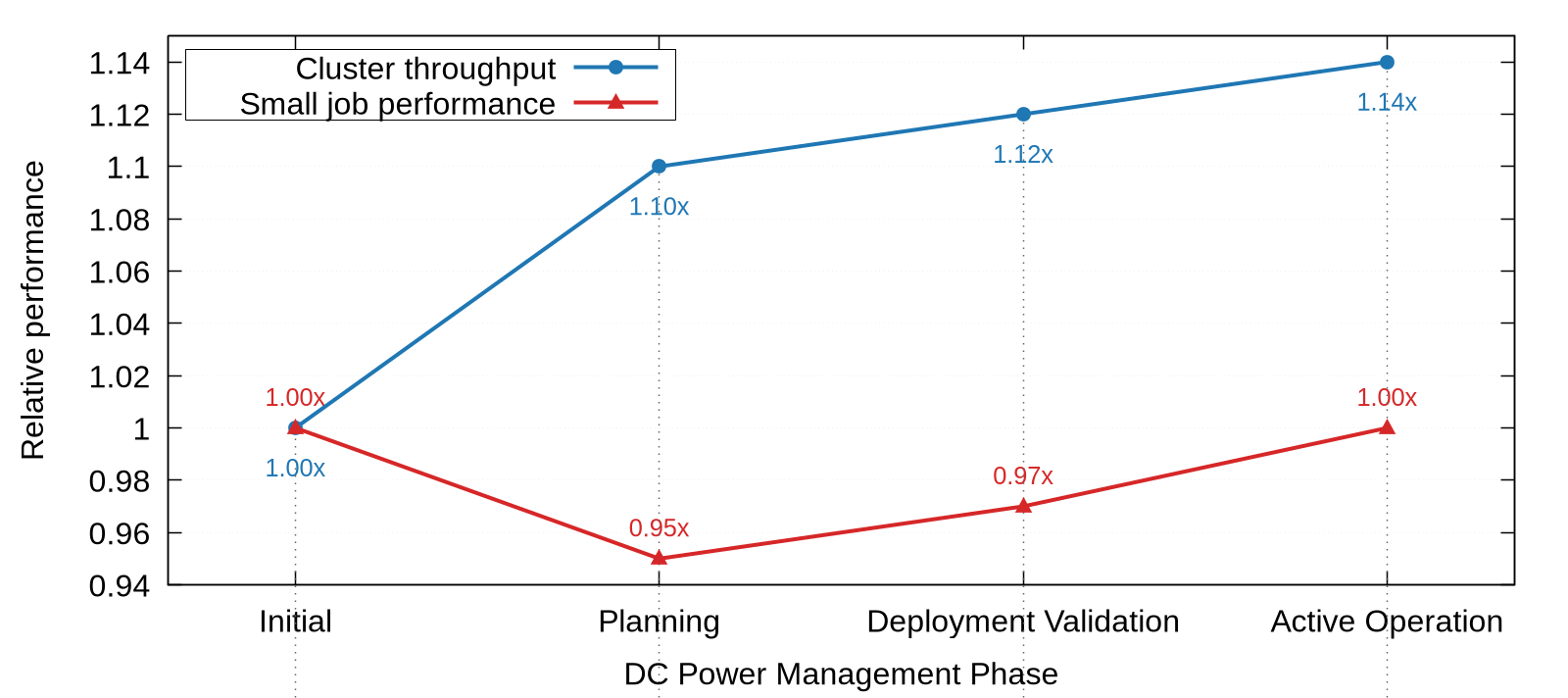}
    \caption{Relative Cluster throughput through Power Management phases. The Figure also presents relative performance of a small job to also show the trade off of per job performance (latency).}
    \label{fig:alltogether}
\end{figure}

Figure~\ref{fig:alltogether} shows the improvement in cluster throughput as we progress through the presented power management phases. We compare the performance against the initial state with 1200W power limit per GPU. The figure shows ~10\% improvement of throughput by optimizing for perf/watt and hence increasing the number of GPUs landed in the DC with 960W power limit per GPU. Then in the deployment validation, we further increase the TDP from initial 960W to 1020W, capturing around 2\% throughput upside. Last, through Dimmer, we
capture the last few percentage of stranded power for further performance increased performance, i.e., extra $\approx$ 2\% upside for the cluster throughput.

\section{Research Wishlist for the Community}
We conclude the paper with two areas that we believe can benefit from further community research.

\noindent \textbf{Workload-aware SoC power sharing}. With GPU, CPU, HBM, LPDDR, and NVLink IPs under a module-level power limit, there is an opportunity to properly manage the shared power across such components to maximize performance, considering different level of sensitivity of end to end performance to each component's power. (Figure~\ref{fig:gemm_tdp}-~\ref{fig:cluster_throughput_tdp}).  We encourage research into microarchitectural mechanisms for autonomous intra-SoC power rebalancing.

\noindent \textbf{Formally power-optimal job scheduling}. Our scheduler optimizes placement based on network topology alone, but the power delivery hierarchy is a separate, overlapping tree. A scheduler that considers both could balance power load across devices while maintaining network locality. This multi-objective trade-off  can be formalized as a mixed-integer nonlinear program (MINLP): maximize aggregate throughput (sum of per-job min-host throughputs) subject to hierarchical power capacity constraints (RPP, SB, MSB). The coupling between placement and power variables, the min operator capturing the straggler effect, and tree-structured constraints make this computationally hard but amenable to decomposition and approximation.

\section{Related Work}

\noindent \textbf{Datacenter Power Management}
Prior work has explored datacenter-wide power management strategies to improve utilization and efficiency. For example, Piga et al.~\cite{piga24} propose safe DVFS boosting to increase capacity, while Flex~\cite{flexDatacenter} and SmartOClock~\cite{smartoclock} leverage workload-aware oversubscription to maximize power budgets. SmoothOperator~\cite{smoothOperator} reduces peak power draw by coordinating the timing of synchronous services. Many systems employ statistical multiplexing across servers~\cite{ensemble1, ensemble2, mvpp, dynamoFB, thunderbolt, predictionVM, needSpeed, distributedUps, smoothOperator} to safely oversubscribe power and reduce costs, with such techniques now deployed in production environments~\cite{awsPstates, azureVMSeries, gcpBoost}. 
These approaches typically require tight cluster-level coordination and deep integration with the software stack.

In contrast, our work addresses both the provisioning and operational phases of power management in large-scale AI training clusters. We focus on practical strategies for ensuring safety, maximizing performance, and adapting to dynamic workload and infrastructure conditions at scale.

\vspace{2pt}
\noindent \textbf{Energy-efficiency in Datacenters}
Prior work has focused on improving energy efficiency for CPU-centric datacenter workloads~\cite{ecofaas, twig, eetl, pegasus}, with more recent research examining the energy characteristics of GPU-based applications~\cite{e_energy, cloudEnergy, wordstowatts, dynamollm}. Recent schemes manage energy and power consumption for DNN inference and training~\cite{zeus, alert, dynamictraining} through techniques such as frequency scaling~\cite{batchDVFS, geepfas, batchsizer, heterDVFS, edgebert, llmDVFS, indicator_directed}, autoscaling~\cite{autoScale}, and resource partitioning and mapping~\cite{wattwiser, edgebert}. These approaches often target fine-grained control at the device or job level, optimizing for throughput or energy proportionality in DNN workloads.

In contrast,  
% that has advanced energy management for individual jobs or devices, 
our work addresses the challenges of power management at the scale of entire AI training clusters. 
% We focus on both provisioning-time and operational strategies that ensure safety, maximize performance, and adapt to dynamic workload and infrastructure conditions in large-scale deployments.

\section{Conclusion}

The explosive growth of AI training workloads has made power availability the primary constraint on further datacenter scaling and infrastructure expansion. 
This paper presented our company's approach to power provisioning and operational management for large-scale AI training clusters. 
By decoupling provisioned and operational rack power, and leveraging detailed measurement across the power hierarchy, we identified key inefficiencies and developed strategies to maximize datacenter-level \emph{\pw}.

Our findings highlight the importance of holistic, system-level power management for sustainable AI infrastructure. 
We hope these insights will guide future efforts in designing efficient, scalable AI datacenters.
% to meet the growing demands of AI workloads.

%-------------------------------------------------------------------------------
\bibliographystyle{plain}
\bibliography{references.bib}

%%%%%%%%%%%%%%%%%%%%%%%%%%%%%%%%%%%%%%%%%%%%%%%%%%%%%%%%%%%%%%%%%%%%%%%%%%%%%%%%
\end{document}